\documentclass[aip,graphicx, 11pt]{revtex4}

\pdfoutput=1

\usepackage{color}
\usepackage{graphics}
\usepackage[pdftex]{graphicx}
\usepackage{amsmath}
\usepackage{amssymb}
\usepackage{amsfonts}
\usepackage{float}
\usepackage{natbib}
\usepackage{verbatim}
\usepackage{algorithm}
\usepackage{algorithmic}
\usepackage{todonotes}
\usepackage{lineno}
\usepackage{bbold}
\usepackage{array}
\usepackage{mathrsfs}
\usepackage{amsbsy}
\usepackage{mathtools}
\usepackage{cleveref}
\usepackage{placeins}
\usepackage{xcolor}
\usepackage[normalem]{ulem}
\usepackage{titlesec}
\titlespacing*{\section}{0pt}{1.0\baselineskip}{\baselineskip}

\newcolumntype{?}{!{\vrule width 1pt}}
\newcolumntype{C}{>{\centering\arraybackslash}p{3.7cm}}
\newcolumntype{L}{>{\raggedright\arraybackslash}p{3.9cm}}
\newcolumntype{e}{>{\raggedright\arraybackslash}p{2.5cm}}
\newcolumntype{k}{>{\raggedright\arraybackslash}p{2cm}}
\newcolumntype{l}{>{\raggedright\arraybackslash}p{3.2cm}}
\newcolumntype{R}{>{\raggedleft\arraybackslash}p{3.7cm}}

\newcommand{\circled}[1]{\Large \textcircled{{\normalsize #1}}}
\newcommand{\rom}[1]{\uppercase\expandafter{\romannumeral #1\relax}}

\begin{document}

\title{Shape curvature effects in viscous streaming}

\author{Yashraj Bhosale}
\affiliation{Mechanical Sciences and Engineering, University of Illinois at Urbana-Champaign, Urbana, IL 61801, USA}

\author{Tejaswin Parthasarathy}
\affiliation{Mechanical Sciences and Engineering, University of Illinois at Urbana-Champaign, Urbana, IL 61801, USA}

\author{Mattia Gazzola}
\email{mgazzola@illinois.edu}
\affiliation{Mechanical Sciences and Engineering and National Center for Supercomputing Applications, University of Illinois at Urbana-Champaign, Urbana, IL 61801, USA}

\begin{abstract}
Viscous streaming flows generated by objects of constant curvature (circular cylinders,
infinite plates) have been well understood. Yet, characterization and 
understanding of such flows when multiple body length-scales are involved has not been 
looked into, in rigorous detail. We propose a simplified setting to understand and 
explore the effect of multiple body curvatures on streaming flows, analysing the system
through the lens of bifurcation theory. 
Our setup consists of periodic, regular lattices of cylinders characterized by two
distinct radii, so as to inject discrete curvatures into the system, which in turn affect
the streaming field generated due to an oscillatory background flow. We demonstrate that 
our understanding based on this system can be then
generalised to a variety of individual convex shapes presenting a spectrum of curvatures, explaining prior
experimental and computational observations. Thus, this study illustrates a route towards the rational
manipulation of viscous streaming flow topology, through regulated variation of object geometry.
\end{abstract}
\maketitle

\vspace{-5pt}
\section{Introduction}\label{sec:intro}
\vspace{-5pt}


This paper investigates the role of body curvature in two-dimensional viscous
streaming phenomena. Viscous streaming refers to the time-averaged steady flow that
arises when an immersed body of characteristic length scale $D$ undergoes
small-amplitude oscillations (compared to $D$) in a viscous fluid. 
Viscous streaming has been well explored and characterized theoretically,
experimentally and computationally, for constant curvature shapes which include oscillating individual 
circular cylinders \citep{holtsmark1954boundary, riley2001steady, lutz2005microscopic, 
coenen2013oscillatory, vishwanathan2019steady}, infinite flat plates
\citep{glauert1956laminar, yoshizawa1974steady} and spheres \citep{Lane:1955, Riley:1966, Kotas:2007}. 
However, little is known beyond these
simple objects, in particular when multiple curvatures in complex shapes are involved.
Efforts have been made in this direction by considering individual oscillating ellipses
\citep{badr1994oscillating}, spheroids \citep{Kotas:2007}, triangle and square cylinders 
\citep{tatsuno1974circulatory, tatsuno1975circulatory}, sharp edges 
\citep{nama2014investigation, ovchinnikov2014acoustic} as well 
as multiple identical cylinders \citep{yan1994streaming, coenen2013oscillatory,
coenen2016steady}. Yet, our understanding of how streaming flow features
and topology are affected by multiple body length scales remains largely incomplete.

Our motivation to understand these relations stems from the broad range of applications of
viscous streaming in
microfluidic flow manipulation, particle trapping, scalar transport and passive swimming
\citep{liu2002bubble, lutz2003microfluidics, marmottant2004bubble,
    nair2007hydrodynamically, chung20093, tchieu2010fluid,
wang2011size, chong2013inertial,klotsa2015propulsion, thameem2016particle, thameem2017fast} which can benefit from an
expanded flow design space based on geometrical variations. Additionally, we are motivated 
by the emergence of artificial and biohybrid mini-bots operating in fluids
\citep{williams2014self, park2016phototactic, ceylan2017mobile, aydin2019neuromuscular, 
huang2019adaptive}. 
Indeed, these bots operate across flow regimes 
where streaming effects
can be important, and may be usefully leveraged, opening new opportunities for
micro-robotics in manufacturing or medicine \citep{ceylan2017mobile}. 
For example, in a recent study \citep{parthasarathy2019streaming}, we showed that
streaming can enhance the contactless transport of passive inertial particles (drug payload) by moving 
cylindrical mini-bots.  There, we also highlighted that morphing a circular cylinder to a suitably 
sculpted shape that combines asymmetry and high rear curvature, can further improve transport. 
We attributed this enhancement to a favourable re-arrangement of the streaming 
flow topology.
This raises the question---how do changes in geometry
from a circular cylinder translate into streaming flow topology organization? 
Can we rigorously predict and manipulate topological transitions through shape
variations for flow design purposes? 

In this work, we attempt to answer these questions by first understanding and
characterizing streaming flow topology in a simplified setting in which circular cylinders
of different radii (i.e. curvatures) are arranged in periodic, regular lattices. This allows us
to inject multiple curvatures in a discrete fashion into our system, enabling a systematic study of
their effects. We analyse the different flow topologies
that arise as we vary the cylinders' curvature ratios and the frequency 
of the background oscillatory flow, and characterize their transitions via bifurcation theory.
 Finally, we demonstrate that our understanding 
can be extended to generalised, individual bodies, aided by comparison
against prior experiments \citep{tatsuno1974circulatory, tatsuno1975circulatory}
and computations \citep{parthasarathy2019streaming}.
Overall, this study elucidates the mechanisms at play when streaming flow topology is
manipulated via regulated variations of shape geometry, thus providing a rational design
approach and physical intuition. 

The work is organized as follows: governing equations and numerical method are recapped in
\S \ref{sec:num}; streaming physics is described in \S \ref{sec:setup};
lattice setup, investigation of different flow topologies and corresponding
transitions are presented in \S \ref{sec:bif}; extension to the design
of arbitrary geometries and comparison against experiments is discussed in 
\S \ref{sec:des}; finally, our findings are summarized and discussed in \S \ref{sec:con}.

\vspace{-5pt}
\section{Governing equations and numerical method}\label{sec:num}
\vspace{-5pt}

We briefly recap the governing equations and the numerical solution technique. We consider
incompressible viscous flows in a periodic or unbounded domain $\Sigma$. In this fluid domain,
immersed solid bodies perform simple harmonic oscillations. The bodies are
density-matched and have support $\Omega$ and boundary $\partial \Omega$ respectively. The 
flow can then be described using the incompressible Navier--Stokes 
equations (\ref{ns})
\begin{equation}
    \nabla \cdot \mathbf{u} = 0; \hspace{0.2cm} \frac{\partial \mathbf{u}}{\partial t} + (
    \mathbf{u} \cdot \nabla) \mathbf{u} = - \frac{\nabla P}{\rho} + \nu \nabla^2
    \mathbf{u}, \; \mathbf{x} \in \Sigma \setminus \Omega
  \label{ns}
\end{equation}
where $\rho$, $P$, $\mathbf{u}$ and $\nu$ are the fluid density, pressure, velocity and
kinematic viscosity, respectively. The dynamics of the fluid--solid system are coupled via
the no-slip boundary condition $\mathbf{u} = \mathbf{u_s}$, where
$\mathbf{u_s}$ is the solid body velocity. The system of equations is then solved using a 
velocity--vorticity formulation with a combination of remeshed vortex methods and 
Brinkmann penalization \citep{Gazzola:2011a}. 
This method has been validated across a range of flow--structure interaction problems, from 
flow past bluff bodies to biological swimming \citep{Gazzola:2011a, gazzola2012flow, 
gazzola2012c, gazzola2014reinforcement, gazzola2016learning}. Recently, it has also been 
shown to effectively capture spatio--temporal scales related to viscous streaming
\citep{parthasarathy2019streaming}. 

\section{Streaming: physics and flow topology}\label{sec:setup}
\vspace{-5pt}
\subsection{Streaming physics: classical case of a circular cylinder}\label{sec:strc}
\vspace{-5pt}

\begin{figure}[h!]
    \centerline{\includegraphics[width=\textwidth]{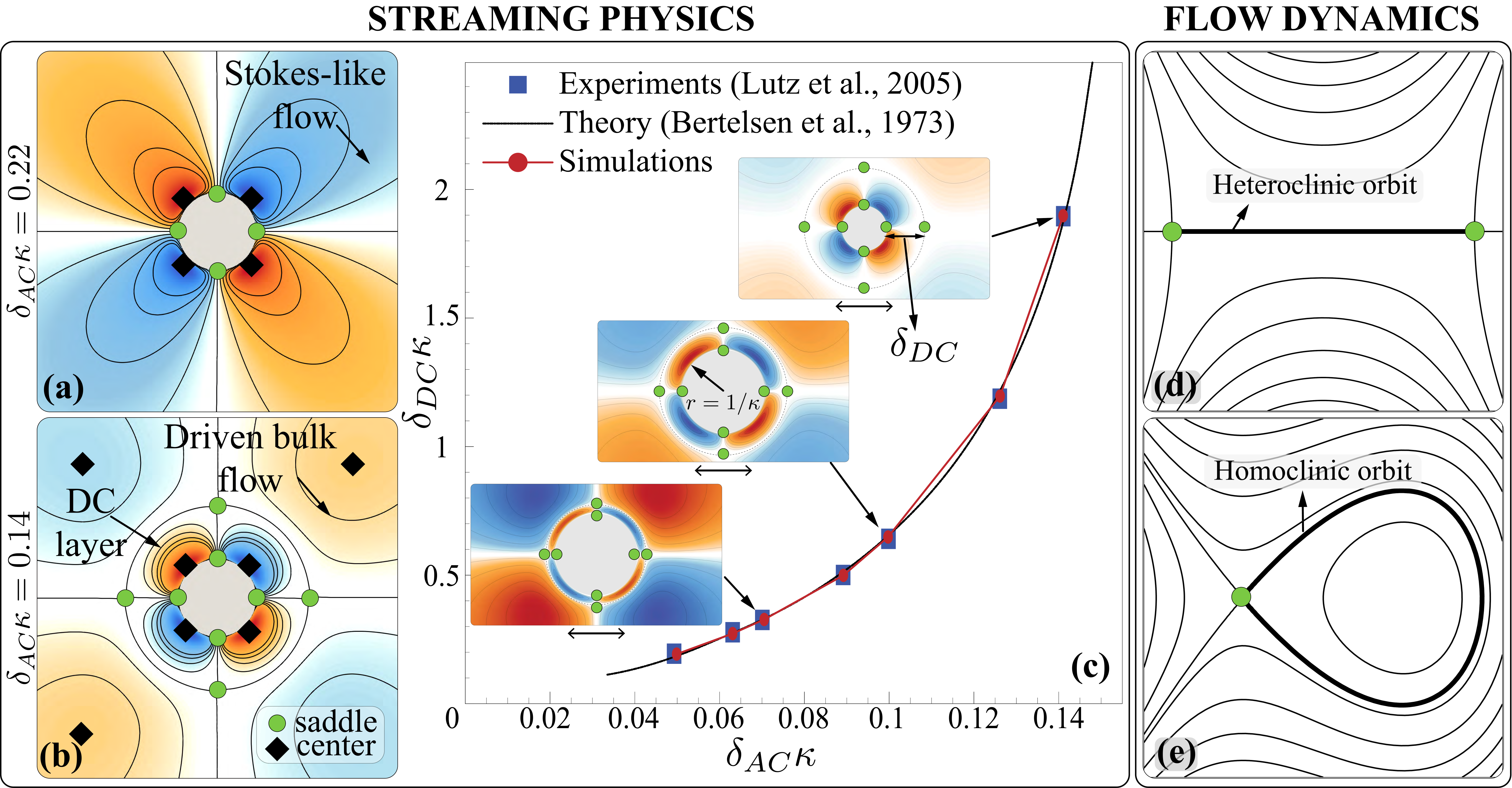}}
    \caption{Streaming characterization in classical circular cylinder setting. 
        Comparison of time-averaged streamline patterns 
        in (\textit{a}) Stokes-like ($\delta_{AC} \kappa = 0.22$) and (\textit{b}) finite-thickness DC layer
        ($\delta_{AC} \kappa = 0.14$) regimes,
        respectively, with the corresponding critical points. Centres and
        saddles (half-saddles on solid boundaries) are denoted as black diamonds and green
        circles, respectively. (\textit{c}) Comparison of normalized DC boundary layer 
        thickness $\delta_{DC} \kappa$ vs. normalized AC boundary layer thickness 
        $\delta_{AC} \kappa$ of our simulations (red)
        against experiments (blue, Lutz \textit{et al.} 2005) and 
        theory (black, Bertelsen \textit{et al.} 1973) in the finite DC layer thickness regime.
        Flow topology: Illustrations showing (\textit{d}) a heteroclinic orbit and
        (\textit{e}) a homoclinic orbit.
        Simulation details: domain $[0,1]^2$ $\textrm{m}^2$, uniform grid spacing $h = 1/2048
        \ \textrm{m}$, penalization factor $\lambda = 10^4$, mollification length $.pdfilon_{moll} =
        2 \sqrt{2} h$, lagrangian CFL $= 0.01$, with viscosity $\nu$ and oscillation frequency
        $\omega$ set according to prescribed streaming Reynolds number ($R_s$). The above values 
        are used throughout the text, unless stated otherwise. We refer to Gazzola
        \textit{et al.} (2011) 
        for details on these parameters.}\label{fig:rec}
\end{figure}

We first characterize streaming in the simple, classical setting of a circular cylinder
undergoing oscillations. We consider a cylinder of constant curvature $\kappa$
(radius $r = 1/\kappa$), in quiescent flow, with an imposed small amplitude oscillatory motion 
$x(t) =  x(0)  + A \sin(\omega t)$ where $A$ and $\omega$ are the dimensional amplitude and 
the angular frequency,
respectively. These small amplitude oscillations ($A \kappa \ll 1$) generate a 
Stokes layer of thickness $\delta_{AC} \sim O(\sqrt{\nu / \omega})$ around 
the cylinder, also known as the AC boundary
layer. The velocity that persists at the edge of this AC layer then drives a viscous 
streaming response in the surrounding fluid \citep{batchelor2000introduction}. 
This streaming response is depicted in
figure \ref{fig:rec}(\textit{a, b}) as clockwise (blue) and anti-clockwise (orange) vortical flow structures 
around the cylinder. We characterize these flow structures using the streaming
Reynolds number $R_s = A^2 \omega / \nu$ \citep{stuart1966double, riley2001steady}. 
Figure \ref{fig:rec}(\textit{a}) shows a flow representative of
$R_s \ll 1$. Such low $R_s$ indicates dominant viscous effects, and indeed the steady streaming
flow is Stokes-like, with characteristic slow velocity decay and recirculating regions extending 
practically to infinity. Figure \ref{fig:rec}(\textit{b}) is representative of 
larger $R_s$ $\sim$ $O(1)-O(10)$, where the interplay of inertial and 
viscous effects leads to the formation of a well defined boundary layer 
of thickness $\delta_{DC}$, also known as the DC boundary layer, which drives the 
fluid in the bulk. 
The normalized DC layer thickness $\delta_{DC} \kappa$  and the AC
layer thickness $\delta_{AC} \kappa$,
can be directly related as illustrated in figure \ref{fig:rec}(\textit{c}) 
\citep{bertelsen1973nonlinear, lutz2005microscopic}. Then, in the
classical constant curvature setting of a single cylinder, specifying 
$\delta_{AC} \kappa$ is sufficient to characterize the streaming flow field and its
topology. This picture breaks down when more complex shapes are considered, and a more generic 
approach to characterize streaming flows becomes necessary. 

\vspace{-10pt}
\subsection{Streaming flow topology: a dynamical systems view}\label{sec:dynt}
\vspace{-5pt}

We propose to characterize the streaming flow topologies generated by complex shape bodies via 
dynamical systems theory. First, we identify critical points in the flow
field, i.e. points where the velocity is zero. These points offer 
a sparse yet complete representation of the flow field and its underlying dynamics
\citep{perry1987description}. Critical points can be classified into saddles and
centres (depending on the local flow properties i.e, eigenvalues of the associated Jacobian), 
and the appearance and disappearance of their connecting streamlines shape the flow and its transitions. 
Figure \ref{fig:rec}(\textit{d}, \textit{e}) illustrates two cases of importance in our
context: heteroclinic orbits defined as streamlines connecting two saddles, and homoclinic
orbits defined as streamlines that connect a saddle to itself, thus forming an enclosed
flow region.
Parametric changes (shape symmetry, body curvature, background flow) lead to the 
displacement of critical points, which can cause the breaking, merging or collapsing of
these orbits, and a consequent topological rearrangement. 

As an illustrative example, we consider again the classical case of a single circular cylinder.
Figure \ref{fig:rec}(\textit{a, b}) depicts the critical points in the
streaming flow field for Stokes-like and finite-thickness DC layer regimes, 
respectively. For reference, centres (vorticity-dominated) are denoted as 
diamonds and saddles (shear-dominated) as circles in figure 
\ref{fig:rec}(\textit{a-e}). Compared to the 
Stokes-like regime, the finite-thickness DC layer regime presents four
additional saddles (on the horizontal and vertical axes), that lie
at a distance $\delta_{DC}$ from the cylinder surface (figure \ref{fig:rec}(\textit{c})). 
Heteroclinic orbits between these saddles form a continuous circular streamline that cleanly
separates the DC layer from the driven fluid, thus helping us to topologically distinguish
the flows. 

The above characterization allows us to investigate flow topology transitions using
bifurcation theory. Since the two dimensional streaming flow in our setting is
time-independent (streamlines $\equiv$ pathlines) and incompressible 
(i.e a streamfunction exists), our system can be
equivalently represented as an autonomous Hamiltonian system with $H \equiv \Psi$, where
$H$ and $\Psi$ correspond to the Hamiltonian and time-averaged streamfunction,
respectively \citep{dam2017topological}.
Due to the $H \equiv \Psi$ equivalence, orbits of streaming fluid particles (iso-contours
of $\Psi$) can be 
interpreted as iso-contours of $H$, enabling us to describe the local flow topology 
using the scalar function $H(x, y)$ alone (which is conserved along a streamline or fluid
orbit).
We exploit this equivalence to map the transitions seen in our lattice system (\S
\ref{sec:bif}) to well-studied bifurcations in Hamiltonian systems. 
Once such a bifurcation is identified, we borrow the corresponding reduced Hamiltonian
form $H(x, y)$, which mathematically captures topology changes near bifurcating critical
points \citep{bosschaert2013bifurcations, strogatz2018nonlinear}. This allows us to predict how
the flow evolves upon perturbing shape curvature and/or background flow conditions.
Moreover, the analysis of the reduced Hamiltonian form provides insight into the
physical mechanisms at play, and guides our intuition of how to manipulate these systems.

\section{Lattice system: setup, phase space and flow bifurcations}\label{sec:bif}

\subsection{Curvature variation setup: cylinders in an infinite, regular lattice}

\begin{figure}[h!]
    \centerline{\includegraphics[width=0.7\textwidth]{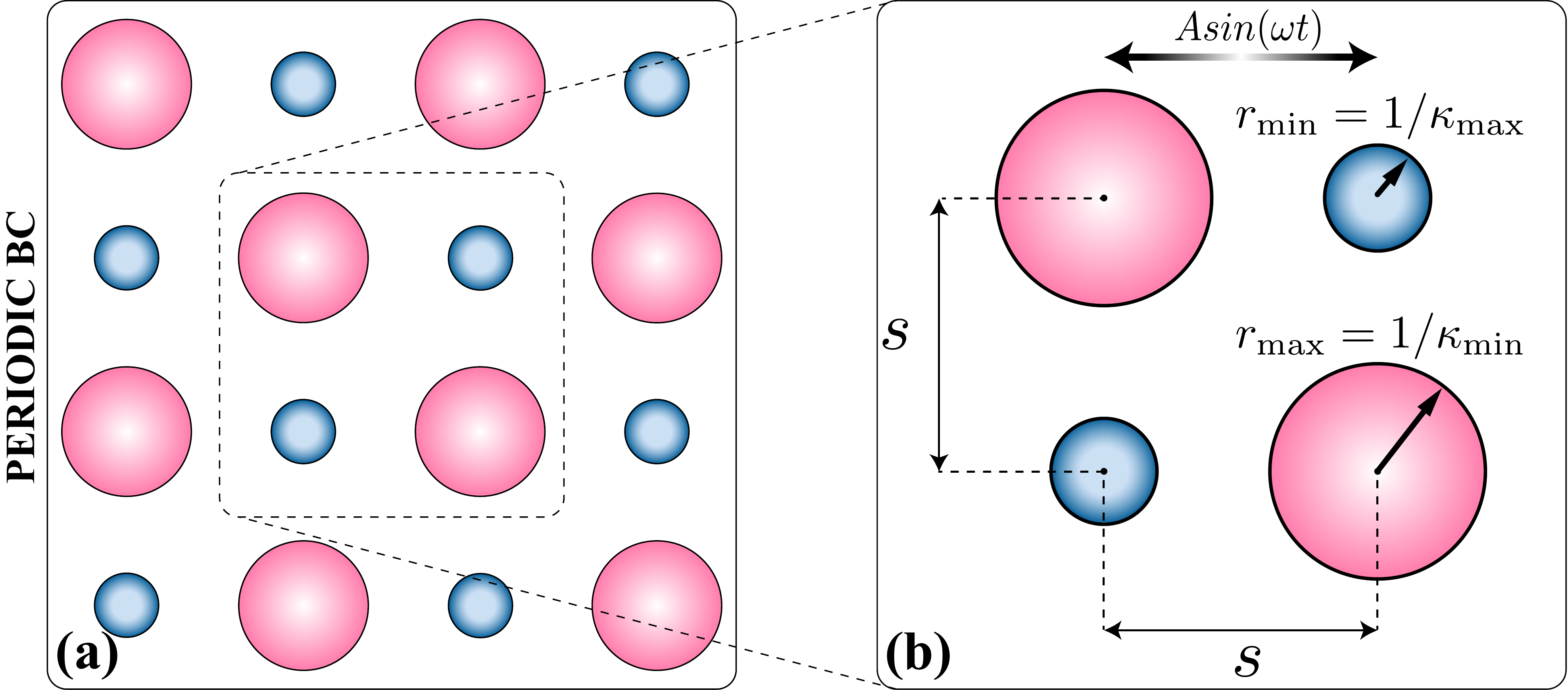}}
    \caption{Curvature variation setup. Illustrations of (\textit{a}) computational domain 
        and regular lattice with periodic boundary conditions. (\textit{b}) A repeating
    unit cell of the lattice system with cylinders of two curvatures
    $\kappa_{\textrm{\scriptsize{max}}}$ and $\kappa_{\textrm{\scriptsize{min}}}$ and the fixed
centre-to-centre spacing $s = 12.5 /  \kappa_{\textrm{\scriptsize{max}}}$.}\label{fig:lat}
\end{figure}

We systematically study body curvature effects via a system consisting of staggered circular 
cylinders of two radii, $1 / \kappa_{\textrm{\scriptsize{max}}}$ and 
$1 / \kappa_{\textrm{\scriptsize{min}}}$, assembled
into a periodic regular lattice (figure \ref{fig:lat}(\textit{a})), with 
$\kappa_{\textrm{\scriptsize{max}}}$ kept constant as a 
reference length scale.
Throughout the study, the centre-to-centre distance $s$ between these cylinders is 
kept constant as $12.5 /  \kappa_{\textrm{\scriptsize{max}}}$, which allows us to vary the
curvature ratio ($\kappa_{\textrm{\scriptsize{max}}} / \kappa_{\textrm{\scriptsize{min}}}$)
from 1 to 6. 
We note here that we performed cursory phase space explorations for different values of $s$, 
and observed that the qualitative nature of the emerging streaming fields is preserved, 
although the boundaries between different topological phases (see next sections) shift quantitatively.
The oscillatory amplitude $A$ for all the cylinders in the 
lattice is kept constant ($A \kappa_{\textrm{\scriptsize{max}}} = 0.1$). 

A variation of $\kappa_{\textrm{\scriptsize{max}}} / \kappa_{\textrm{\scriptsize{min}}}$ 
in the system manifests as a variation in the local AC layer thickness 
($\delta_{AC} \kappa_{\textrm{\scriptsize{max}}} = 
A \kappa_{\textrm{\scriptsize{max}}}/ \sqrt{R_s}$ and 
$\delta_{AC} \kappa_{\textrm{\scriptsize{min}}} = 
A \kappa_{\textrm{\scriptsize{min}}} / \sqrt{R_s}$), and thus in the DC layer thickness, 
with both affecting flow topology.
With $\kappa_{\textrm{\scriptsize{max}}} / \kappa_{\textrm{\scriptsize{min}}}$ capturing all
geometric variation, and $\delta_{AC} \kappa_{\textrm{\scriptsize{max}}}$ capturing all
background flow variation (\S \ref{sec:strc}), we set to map the corresponding phase
space.  
We hypothesize that the flow dynamics underlying this two discrete-curvatures 
setup will generalize to individual, complex shapes with a range of curvatures.

\begin{figure}[h!]
    \centerline{\includegraphics[width=\textwidth]{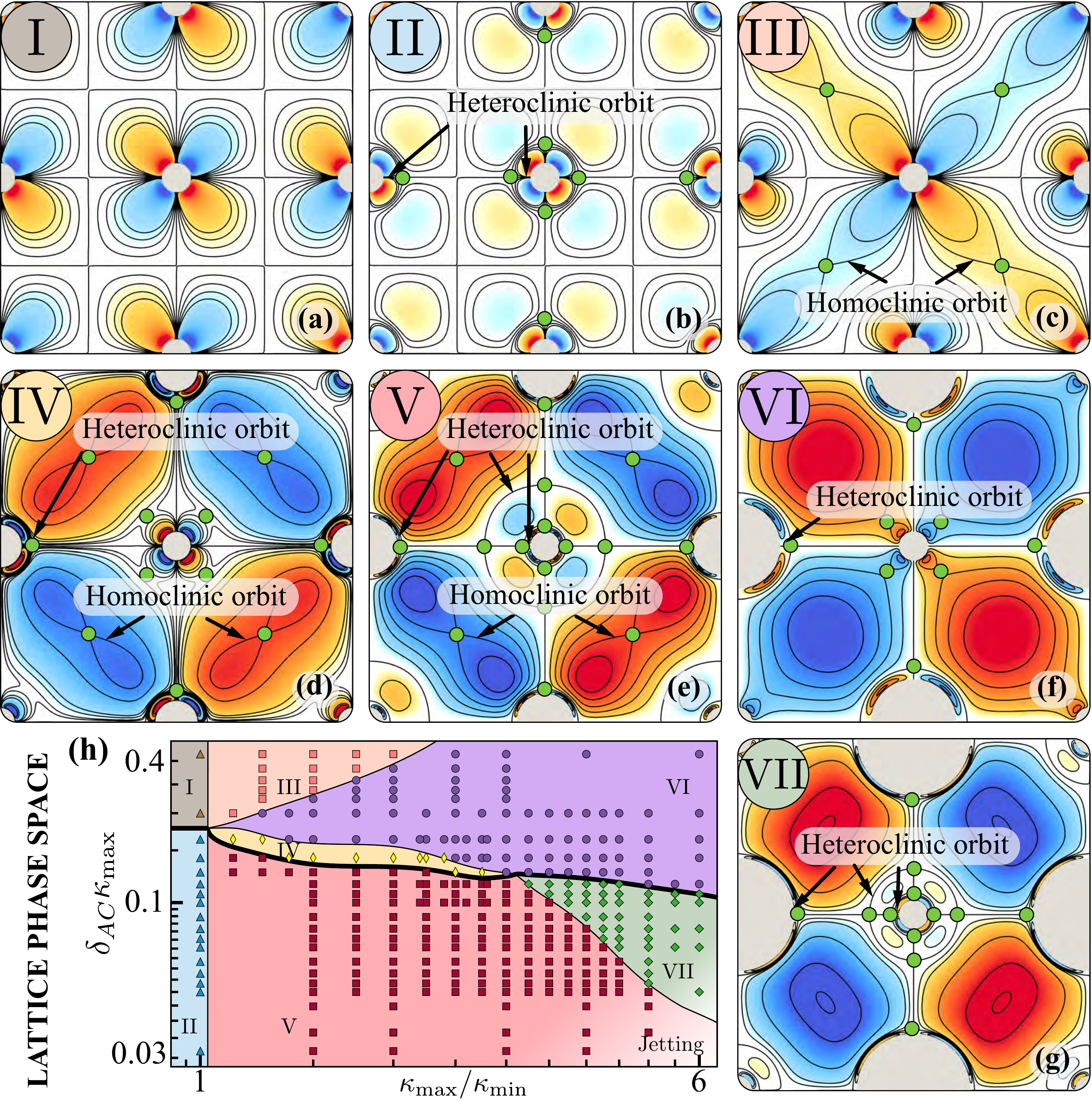}}
    \caption{Lattice phase space. The time-averaged streamline patterns depicted in
        (\textit{a-g}) are classified into different phases depending on their flow
        topology.
        Defining connections and corresponding saddles (green circles) are
    highlighted in (\textit{a-g}). (\textit{h}) Phase space as a function of $\delta_{AC}
\kappa_{\textrm{\scriptsize{max}}}$ and 
$\kappa_{\textrm{\scriptsize{max}}} / \kappa_{\textrm{\scriptsize{min}}}$. Black lines
indicate transition boundaries between phases. The bold line indicates the existence of a
hidden phase, which is characterized later.}\label{fig:ps}
\end{figure}
\vspace{-5pt}
\subsection{Lattice system: phase space}
\vspace{-5pt}
We proceed
with the systematic variation of $\delta_{AC} \kappa_{\textrm{\scriptsize{max}}}$ and
$\kappa_{\textrm{\scriptsize{max}}} / \kappa_{\textrm{\scriptsize{min}}}$,
and span the phase space shown in figure \ref{fig:ps}(\textit{h}). Here, we classify the observed
flow topological patterns into distinct phases, based on critical points and orbits. 
We observe seven main phases.
\vspace{-15pt}
\subsubsection{Phase \rom{1}}
\vspace{-15pt}
Figure \ref{fig:ps}(\textit{a}) shows a representative flow pattern of Phase \rom{1}.
The flow around each cylinder is perfectly repeating due to constant curvature
($\kappa_{\textrm{\scriptsize{max}}} / \kappa_{\textrm{\scriptsize{min}}} = 1$)
and symmetry, and presents only the DC layers around the cylinder. This is a direct
generalization of figure \ref{fig:rec}(\textit{a}) to multiple, identical cylinders.
\vspace{-15pt}
\subsubsection{Phase \rom{2}}
\vspace{-15pt}
The flow is perfectly repeating around each cylinder on account of the constant curvature
($\kappa_{\textrm{\scriptsize{max}}} / \kappa_{\textrm{\scriptsize{min}}} = 1$)
and symmetry, and presents both the driven flow regions (separated by heteroclinic
orbits) and the DC layers around the cylinder (figure \ref{fig:ps}(\textit{b})).
This is a direct generalization of figure \ref{fig:rec}(\textit{b}) to multiple, identical cylinders.
\vspace{-15pt}
\subsubsection{Phase \rom{3}}
\vspace{-15pt}
The DC layers of the smaller cylinders interact with each other, while those of 
the larger cylinders do not (figure \ref{fig:ps}(\textit{c})). This leads to the 
formation of a homoclinic orbit which joins the saddle
at the centre of the unit cell to itself.
\vspace{-15pt}
\subsubsection{Phase \rom{4}}
\vspace{-15pt}
The driven flow regions of the larger cylinders 
interact with each other (while those of the smaller cylinders do not), forming a 
homoclinic orbit which joins the saddle at the centre of the unit cell
to itself (figure \ref{fig:ps}(\textit{d})). Additionally, around the smaller cylinders
only the DC layers are observed.
\vspace{-15pt}
\subsubsection{Phase \rom{5}}
\vspace{-15pt}
The interaction of the driven flow regions of the larger cylinders forms a homoclinic
orbit (figure \ref{fig:ps}(\textit{e})). Additionally, new driven flow regions are
observed around the smaller cylinders. These buffer regions are separated from the smaller
cylinders' DC layers and the homoclinic orbit regions via heteroclinic orbits.
\vspace{-15pt}
\subsubsection{Phase \rom{6}}
\vspace{-15pt}
The driven flow regions of the larger cylinders merge to form a single vortical flow region (no
homoclinic orbit), while only the DC layers are observed around the smaller cylinders
(figure \ref{fig:ps}(\textit{f})).

\vspace{-15pt}
\subsubsection{Phase \rom{7}}
\vspace{-15pt}
Along with the merging of the driven flow regions of larger cylinders 
(no homoclinic orbit), around the smaller cylinders both the buffer driven flow regions 
(separated by heteroclinic orbits) 
and the DC layers are observed (figure \ref{fig:ps}(\textit{g})).
\vspace{-15pt}
\subsubsection{Hidden Phase}\label{sec:hp}
\vspace{-15pt}
Besides the main seven phases reported above,  we encounter a hidden phase along the 
\rom{1} $\to$ \rom{2}, \rom{4} $\to$ \rom{5} and \rom{6} $\to$ \rom{7} boundaries. In our lattice setup, this 
phase is a very narrow sliver characterized by fine flow structures. Since this phase 
would be hardly visible in our phase space, we indicate it as a marked bold line, and 
postpone its characterization when the corresponding flow transitions are analyzed in \S
\ref{sec:umb}. Nonetheless, this phase is important and becomes more prominent when 
shapes other than circular cylinders are considered, as demonstrated in \S \ref{sec:des}.
\vspace{-15pt}
\subsubsection{Jetting Regime}
\vspace{-15pt}
Finally, we note the presence of a jetting regime (bottom-right corner of the phase
space), characterized by unsteady jets developing from the
cylinder surface along the oscillation direction. This phenomenon is well-known
\citep{davidson1972jets, bertelsen1974experimental} and is captured by our
solver \citep{parthasarathy2018viscous}. However, the current study focuses on steady
streaming phenomena and we will not be looking at jetting here. 

\vspace{3mm}
In summary, we observe that curvature variations give rise to rich dynamics. This 
manifests in a variety of flow topologies that are not merely the superposition of 
streaming fields of the individual cylinders (Phase \rom{1} and \rom{2}), but also 
emerge from their non-linear interactions as a collective behavior response.
\vspace{-5pt}
\subsection{Lattice system: flow bifurcations}
\vspace{-5pt}

Next, we characterize the topological transitions between phases from a dynamical
systems perspective, using bifurcation theory. 

\vspace{-10pt}
\subsubsection{Phase \rom{2} $\to$ \rom{5}: heteroclinic orbit bifurcation}\label{sec:sym}
\vspace{-10pt}

\begin{figure}[h!]
    \centerline{\includegraphics[width=\textwidth]{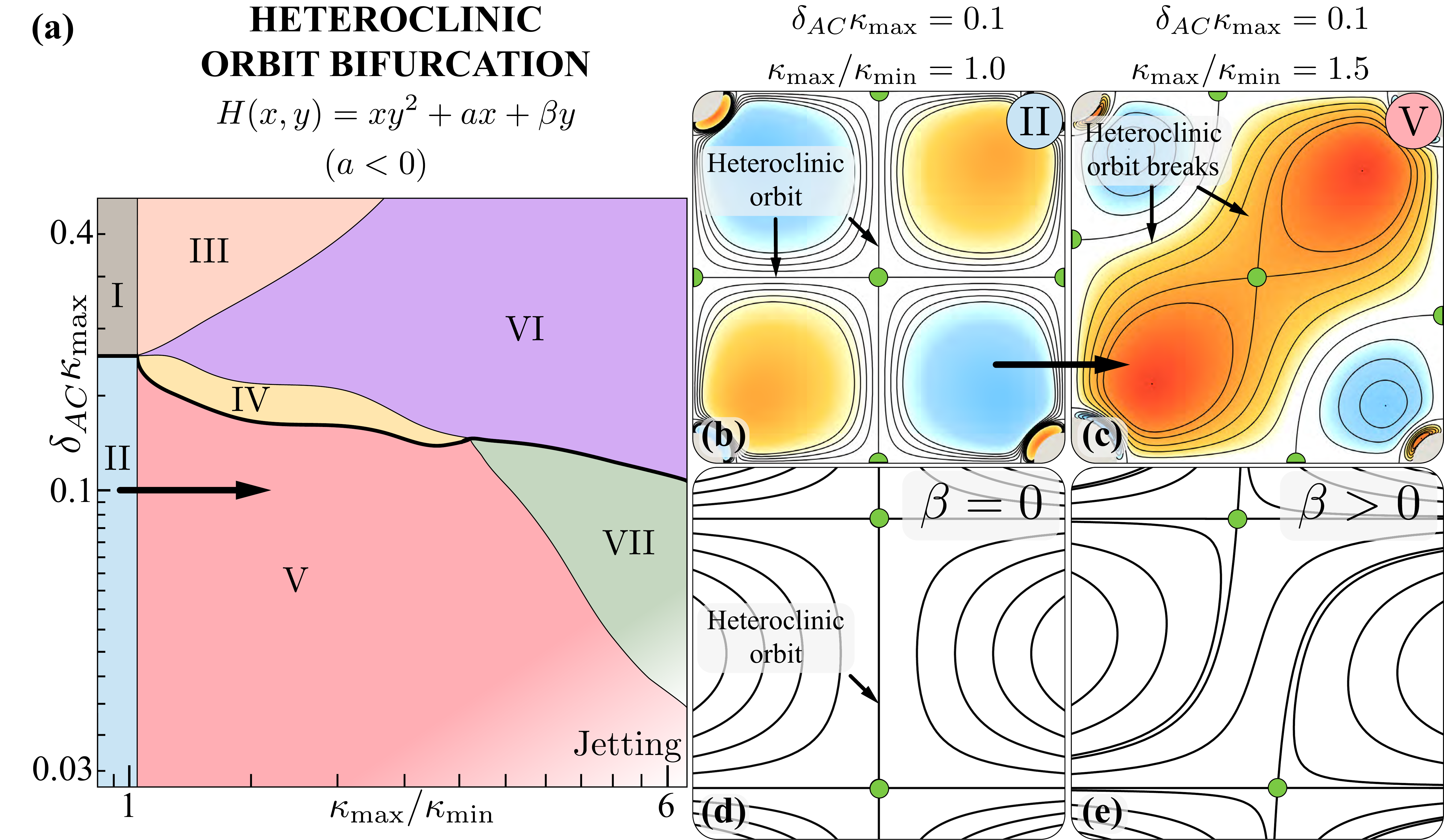}}
    \caption{Phase \rom{2} $\to$ \rom{5}: heteroclinic orbit bifurcation. (\textit{a}) 
        The transition is highlighted on the phase space and the corresponding reduced
        Hamiltonian form is reported. (\textit{b, c}) Flows representative of Phase
        \rom{2} and \rom{5}, respectively. (\textit{d, e}) Bifurcations captured as
        contours of the reduced Hamiltonian form.
   }\label{fig:lat_hetero}
\end{figure}

We first consider the phase transition 
\rom{2} $\to$ \rom{5} in figure \ref{fig:lat_hetero}(\textit{a}). 
We draw attention to the presence of heteroclinic orbits in Phase
\rom{2} (figure \ref{fig:lat_hetero}(\textit{b})) and their absence in
Phase \rom{5} (figure \ref{fig:lat_hetero}(\textit{c})). 
The simplest Hamiltonian form that captures this transition, in terms of critical points,
orbits and symmetry before and after, can be expressed as $H(x, y) = x y^{2} + a x +
\beta y$ with $a < 0$ \citep{kuznetsov2013elements}. 
Here $\beta y$ is the unfolding term, which is added to the mathematical representation of
the dynamical system to investigate its behaviour upon a perturbation
\citep{murdock2006unfoldings}.
As can be seen, if $\beta = 0$ the Hamiltonian is perfectly symmetric about the horizontal
and vertical axes
($x = y = 0$, located in the middle of figure \ref{fig:lat_hetero}(\textit{d})) and exhibits two saddles connected
by a heteroclinic orbit. As the unfolding term deviates from zero ($\beta \neq 0$) the
orbit breaks up (figure \ref{fig:lat_hetero}(\textit{e})) and the system undergoes a
heteroclinic orbit bifurcation \citep{kuznetsov2013elements}, as observed in figure
\ref{fig:lat_hetero}(\textit{c}) and reflected in Phase \rom{5}. The identification of the
bifurcation type in mathematical terms provides insight into the mechanisms at play.
Indeed, $\beta \neq 0$ is physically interpreted as introducing asymmetry in the system,
which we achieve through curvature variation
($\kappa_{\textrm{\scriptsize{max}}} / \kappa_{\textrm{\scriptsize{min}}} > 1$).
Nonetheless, symmetry can be broken in any number of ways, leading to the same flow
topology rearrangement. As a demonstration (shown in the appendix), we
recover the same orbit bifurcation by keeping
$\kappa_{\textrm{\scriptsize{max}}} / \kappa_{\textrm{\scriptsize{min}}} = 1$, while
breaking symmetry via a slow uniform background flow.

This example illustrates how the phase space combined with bifurcation analysis, can
provide a set of rules to understand and manipulate streaming flow topology.

\vspace{-10pt}
\subsubsection{Phase \rom{5} $\to$ \rom{7}: supercritical pitchfork bifurcation}\label{sec:curv}
\vspace{-10pt}

\begin{figure}[h!]
    \centerline{\includegraphics[width=\textwidth]{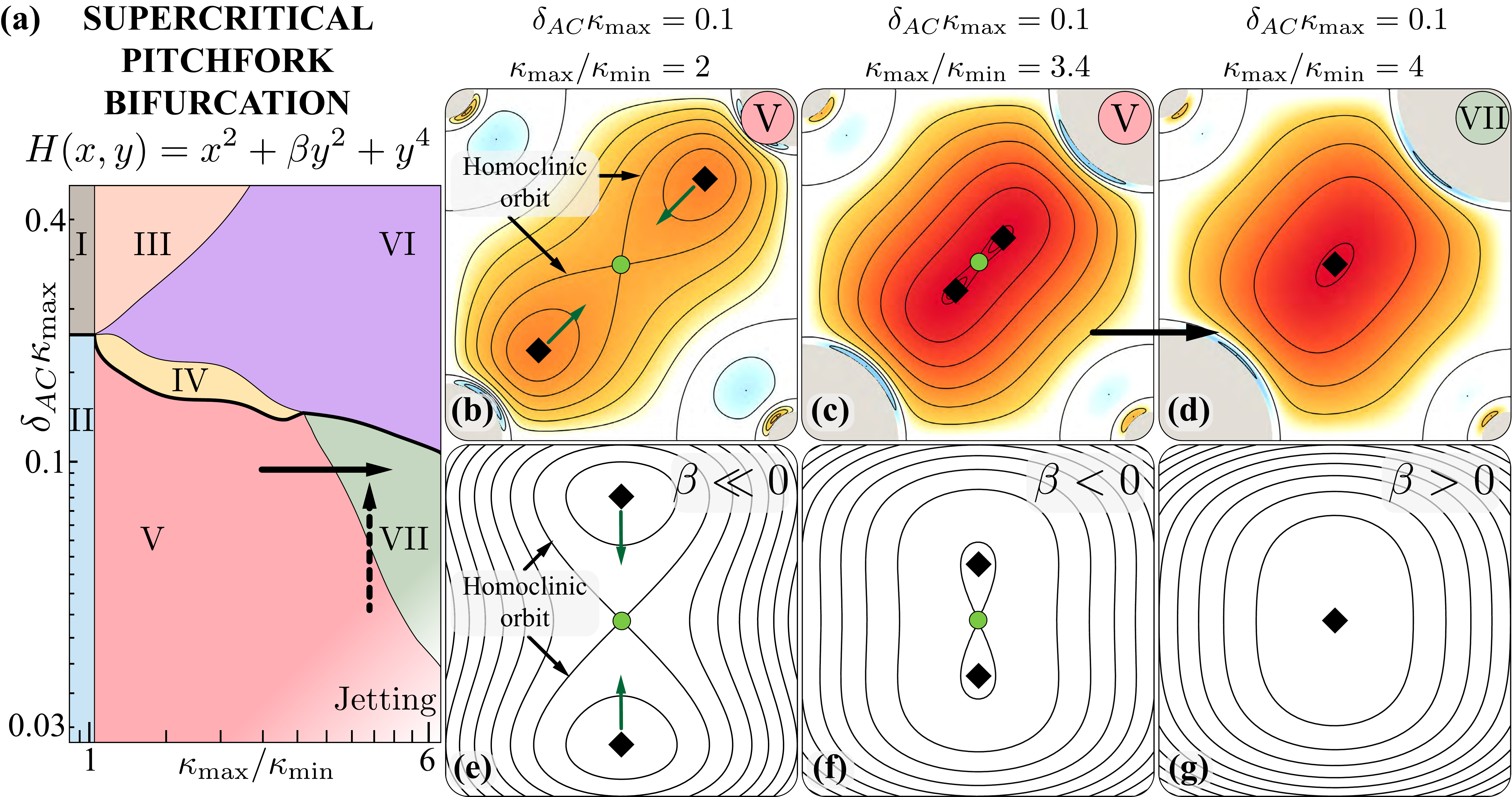}}
    \caption{Phase \rom{5} $\to$ \rom{7}: supercritical pitchfork bifurcation. (\textit{a}) 
        The transition is highlighted on the phase space and the corresponding reduced
        Hamiltonian form is reported. (\textit{b, c, d}) Flows representative of Phase
        \rom{5}, Phase \rom{5} approaching the transition, and Phase \rom{7}, respectively. 
        (\textit{e, f, g}) Bifurcations captured as
        contours of the reduced Hamiltonian form. This bifurcation can also be triggered by
        varying the background oscillatory flow (i.e. by increasing $\delta_{AC}$,
        illustrated  with a dashed
        vertical arrow in (\textit{a})), which is reflected in the phase space as an 
        inclined transition boundary.}
\label{fig:lat_suppitchf}
\end{figure}

We discuss the bifurcation Phase \rom{5} $\to$ \rom{7}, as depicted in 
figure \ref{fig:lat_suppitchf}(\textit{a}).
We draw attention to the presence of homoclinic orbits (with two enclosed centres and a
saddle) in Phase \rom{5} (figure \ref{fig:lat_suppitchf}(\textit{b})) and 
their absence (only one centre) in
Phase \rom{7} (figure \ref{fig:lat_suppitchf}(\textit{d})). 
The simplest Hamiltonian form that captures this transition is 
$H(x, y) = x^{2} + \beta y^{2} + y^{4}$, which corresponds to a supercritical pitchfork
bifurcation \citep{buono2005symmetric}. 
Here $\beta y^{2}$ is the unfolding term, and represents the distance between the centres 
(figure \ref{fig:lat_suppitchf}(\textit{e-g})).
In our lattice system, this distance can be directly controlled by increasing
$\kappa_{\textrm{\scriptsize{max}}} / \kappa_{\textrm{\scriptsize{min}}}$, thus increasing
the radii of the two opposite cylinders so as to push the centres towards the saddle in
the middle of the cell (figure \ref{fig:lat_suppitchf}(\textit{c})), causing them to 
collide and destroy the homoclinic orbits (figure \ref{fig:lat_suppitchf}(\textit{d})). We
observe that an equivalent flow topology rearrangement can be triggered by varying the
background oscillatory flow. Indeed, by reducing the streaming Reynolds number $R_s$ 
(i.e. $\delta_{AC} \uparrow$), we can
increase the thickness $\delta_{DC}$ of the inner boundary layers around the cylinders,
which in turn push the
centres to collide with the saddle. As a consequence, the same supercritical pitchfork
bifurcation is also encountered on increasing $\delta_{AC}$ (dashed vertical arrow), explaining
the fact that the boundary between Phase \rom{5} and \rom{7} is inclined.

\vspace{-10pt}
\subsubsection{Phase \rom{3} $\to$ \rom{6}: subcritical pitchfork
bifurcation}\label{sec:curv2}
\vspace{-10pt}

\begin{figure}[h!]
    \centerline{\includegraphics[width=\textwidth]{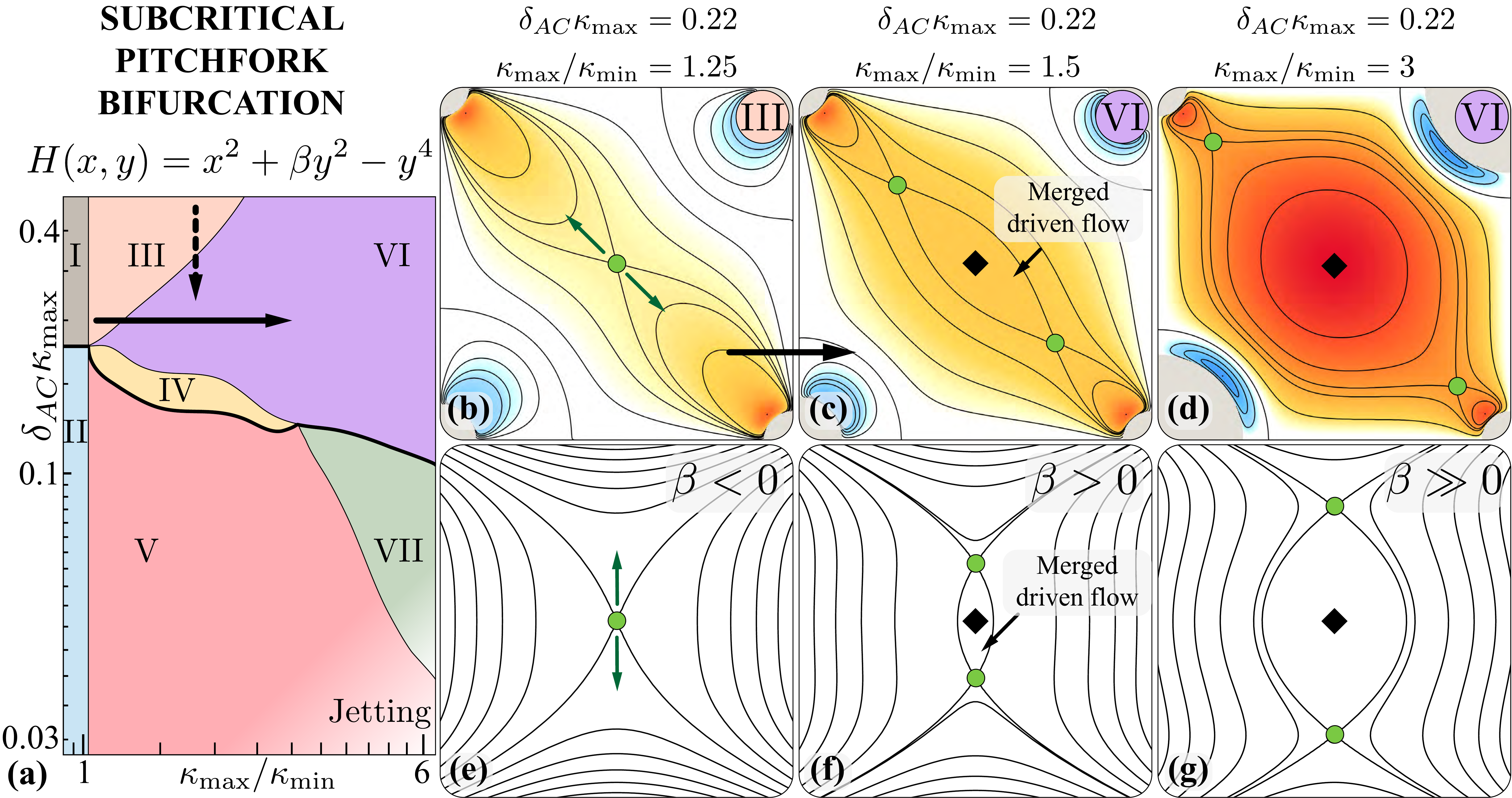}}
    \caption{
        Phase \rom{3} $\to$ \rom{6}: subcritical pitchfork bifurcation. (\textit{a}) 
        The transition is highlighted on the phase space and the corresponding reduced
        Hamiltonian form is reported. (\textit{b, c, d}) Flows representative of Phase
        \rom{3}, Phase \rom{6} approaching the transition, and Phase \rom{6}, respectively. 
        (\textit{e, f, g}) Bifurcations captured as
        contours of the reduced Hamiltonian form. This bifurcation can also be triggered by
        varying the background oscillatory flow (i.e. by increasing $\delta_{AC}$,
        illustrated with a dashed
        vertical arrow in (\textit{a})), which is reflected in the phase space as an 
        inclined transition boundary.
        }\label{fig:lat_subpitchf}
\end{figure}

We now investigate the bifurcation 
Phase \rom{3} $\to$ \rom{6}, as depicted in figure \ref{fig:lat_subpitchf}(\textit{a}).
We draw attention to the absence of a merged driven flow region
in Phase \rom{3} (figure \ref{fig:lat_subpitchf}(\textit{b})) and its
presence (enclosed centre and two saddles) in Phase
\rom{6} (figure \ref{fig:lat_subpitchf}(\textit{d})). 
The simplest Hamiltonian form that captures this transition is 
$H(x, y) = x^{2} + \beta y^{2} - y^{4}$, which corresponds to a subcritical pitchfork 
bifurcation \citep{buono2005symmetric}. 
Here $\beta y^{2}$ is the unfolding term, which drives the appearance of the merged driven flow region ($\beta >
0$) by modulating the distance between the saddles (figure
\ref{fig:lat_subpitchf}(\textit{e-g})).  
In our lattice system, this appearance can be controlled by increasing 
$\kappa_{\textrm{\scriptsize{max}}} / \kappa_{\textrm{\scriptsize{min}}}$, 
which in turn decreases the thickness $\delta_{DC}$ of the inner boundary 
layers around the larger cylinders. This pulls on the saddle in the 
middle (figure \ref{fig:lat_subpitchf}(\textit{b})), 
eventually splitting it into a center and two saddles (figure \ref{fig:lat_subpitchf}(\textit{c})), 
which are further pulled apart as the merged driven
flow grows larger (figure \ref{fig:lat_subpitchf}(\textit{d})). An equivalent flow
topology rearrangement can be achieved by modulating the background flow so as to directly decrease
$\delta_{AC}$ and the thickness $\delta_{DC}$ of the 
inner boundary layers, again pushing the saddles apart and causing the appearance of
a merged driven flow. As a consequence, the same subcritical pitchfork
bifurcation is encountered on decreasing $\delta_{AC}$ (dashed vertical arrow),
explaining the fact that the boundary between Phase \rom{3} and \rom{6} is inclined.

\vspace{-10pt}
\subsubsection{Phase \rom{6} $\to$ \rom{7}: reflecting umbilic bifurcation}\label{sec:umb}
\vspace{-10pt}

\begin{figure}[h!]
    \centerline{\includegraphics[width=0.9\textwidth]{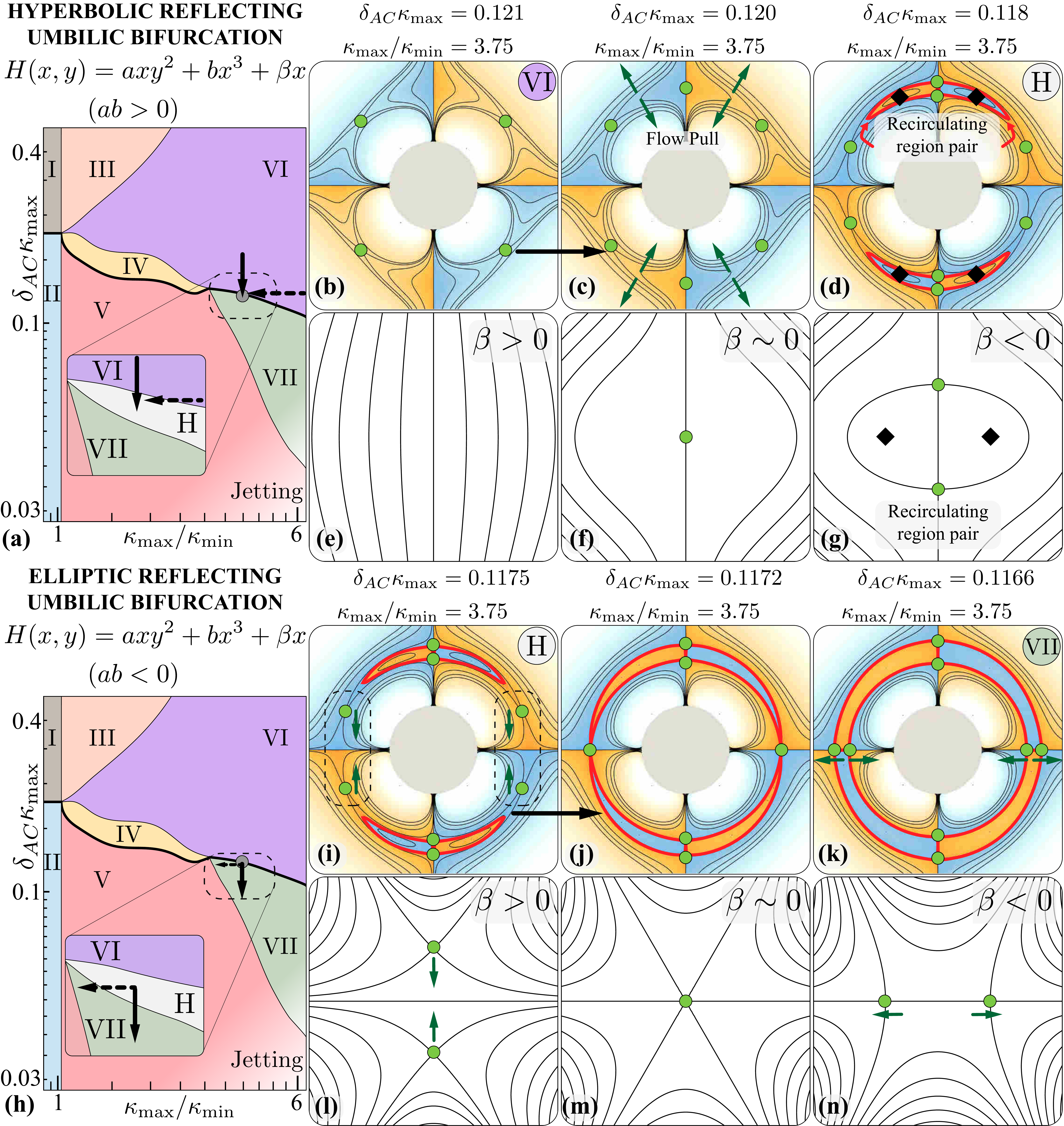}}
    \caption{
        Phase \rom{6} $\to$ hidden Phase H: hyperbolic reflecting umbilic bifurcation. (\textit{a}) 
        The transition is highlighted on the phase space (with a zoomed in view) and the corresponding reduced
        Hamiltonian form is reported. (\textit{b, c, d}) Flows (coloured in a logarithmic
        scale) representative of Phase
        \rom{6}, at the transition, and hidden Phase H, respectively. 
        (\textit{e, f, g}) Bifurcations captured as contours of the reduced Hamiltonian form.        
        (\textit{h}) The transition from hidden Phase H $\to$ \rom{7} (elliptic reflecting
        umbilic bifurcation) is highlighted on the phase space (with a zoomed in view) and the 
        corresponding reduced Hamiltonian form is reported. (\textit{i, j, k}) Flows representative of
        hidden Phase H, at the transition, and Phase \rom{7}, respectively. 
        (\textit{l, m, n}) Bifurcations captured as
        contours of the reduced Hamiltonian form. These bifurcations can also be triggered by
        varying the curvature (i.e by varying 
        $\kappa_{\textrm{\scriptsize{max}}} / \kappa_{\textrm{\scriptsize{min}}}$, a horizontal
        dashed arrow in (\textit{a, h})), which is reflected in the phase space as an inclined
        transition boundary. The newly created recirculating region pairs are marked in
        red. 
        }\label{fig:lat_umbcomb}
\end{figure}

Here, we illustrate the bifurcation Phase \rom{6} $\to$ \rom{7}, as depicted 
in figure \ref{fig:lat_umbcomb}(\textit{a}). 
To identify this bifurcation we focus only on the region local to the smaller cylinder. 
We note the absence of a buffer driven flow region around the smaller cylinder
in Phase \rom{6} (figure \ref{fig:lat_umbcomb}(\textit{b})) and its presence in phase
\rom{7} (figure \ref{fig:lat_umbcomb}(\textit{k}), marked in red).
This flow topology change occurs in two consecutive s.pdf, passing through the hidden phase of \S \ref{sec:hp}. 

In the first step, we draw attention to the absence of recirculating region pairs
in Phase \rom{6} (figure \ref{fig:lat_umbcomb}(\textit{b})) and their presence in
figure \ref{fig:lat_umbcomb}(\textit{d}) (marked in red and comprising two saddles and two centers). 
We note that the latter flow field corresponds to the hidden Phase H.  
The simplest Hamiltonian form that captures this transition is $H(x, y) = a x y^{2} + b x^{3} +
\beta x$ with $a b > 0$, which corresponds to a hyperbolic reflecting umbilic bifurcation 
\citep{bosschaert2013bifurcations}. Here $\beta x$ is the unfolding term, that controls 
the appearance (going from $\beta > 0$ to $\beta < 0$) of the recirculating region pairs
and their size (figure \ref{fig:lat_umbcomb}(\textit{e-g})).  
In our lattice system, the appearance and size of these regions can be
controlled by decreasing $\delta_{AC}$, which decreases the DC layer thickness
$\delta_{DC}$ of both small and large cylinders.
This pulls the streamlines adjacent to the small cylinders's DC layers in two opposite directions 
(figure \ref{fig:lat_umbcomb}(\textit{c})),
forming a degenerate saddle on the vertical axis, which eventually splits into two saddles
and two centres
(figure \ref{fig:lat_umbcomb}(\textit{d})). Topologically, this manifests as a pair of 
counter-rotating recirculating regions.
An equivalent flow topology rearrangement can be alternatively achieved by decreasing
$\kappa_{\textrm{\scriptsize{max}}} / \kappa_{\textrm{\scriptsize{min}}}$, which
increases the distance between the cylinder surfaces. This again pulls the streamlines in
the above described fashion, triggering the same bifurcation.
 As a consequence, 
the same hyperbolic reflecting umbilic bifurcation is also encountered on decreasing
$\kappa_{\textrm{\scriptsize{max}}} / \kappa_{\textrm{\scriptsize{min}}}$ (dashed horizontal 
arrow), explaining the fact that the transition boundary is inclined.

The second step of the Phase \rom{6} $\to$ \rom{7} transition occurs right after further 
decreasing $\delta_{AC}$, rendering the hidden Phase H very narrow.
We focus on the highlighted saddles close to the horizontal
axis in the hidden Phase H (figure \ref{fig:lat_umbcomb}(\textit{i})).
After the transition these saddles are located on the horizontal axis, thus recovering  
Phase \rom{7} (figure \ref{fig:lat_umbcomb}(\textit{k})).
The simplest Hamiltonian form that captures this rearrangement is $H(x, y) = a x y^{2} + b x^{3} +
\beta x$ with $a b < 0$, which corresponds to an elliptic reflecting umbilic bifurcation 
\citep{bosschaert2013bifurcations}. Here $\beta x$ is the unfolding term, that
captures whether the saddles are present ($\beta < 0$) or absent ($\beta > 0$) on the
horizontal axis (figure \ref{fig:lat_umbcomb}(\textit{l-n})), as well as their distance.  
Similar to the previous step, a decrease in $\delta_{AC}$ causes a pull on the streamlines
immediately adjacent to the small cylinder's DC layers. This time though, we do not observe a formation
of degenerate saddles on the horizontal axis. This is due to the
asymmetry resulting from the recirculating region pairs (marked in red in figure
\ref{fig:lat_umbcomb}(\textit{i-k})) generated at the previous step. Instead, the saddles are now pushed towards 
the horizontal midplane, extending the recirculating region pairs. Upon reaching the
midplane, the two opposite saddles collapse (figure \ref{fig:lat_umbcomb}(\textit{j})) and
split along the horizontal axis (figure \ref{fig:lat_umbcomb}(\textit{k})). These new saddles
together with the one formed at the previous step, completely define the buffer driven flow
region around the smaller cylinder.
Again, an equivalent flow topology rearrangement can be achieved by decreasing
$\kappa_{\textrm{\scriptsize{max}}} / \kappa_{\textrm{\scriptsize{min}}}$.
As a consequence, the same elliptic reflecting umbilic bifurcation is also encountered on decreasing
$\kappa_{\textrm{\scriptsize{max}}} / \kappa_{\textrm{\scriptsize{min}}}$ (dashed horizontal 
arrow), explaining the fact that the transition boundary is inclined.

We note here that the order (upon decreasing $\delta_{AC}$) of these transitions is
robust: first the hyperbolic and then the elliptic reflecting umbilic bifurcation.
Nonetheless, the location at which they take place may vary. In our example the
recirculating pairs are formed at the top/bottom of the cylinder, and then extend towards
the horizontal midplane. Alternatively the pairs may form on the left/right of the
cylinder and then grow towards the vertical midplane. More details can be found in the
appendix.

\vspace{-10pt}
\subsection{Summary of bifurcations}
\vspace{-10pt}

\begin{figure}[h!]
    \centerline{\includegraphics[width=\textwidth]{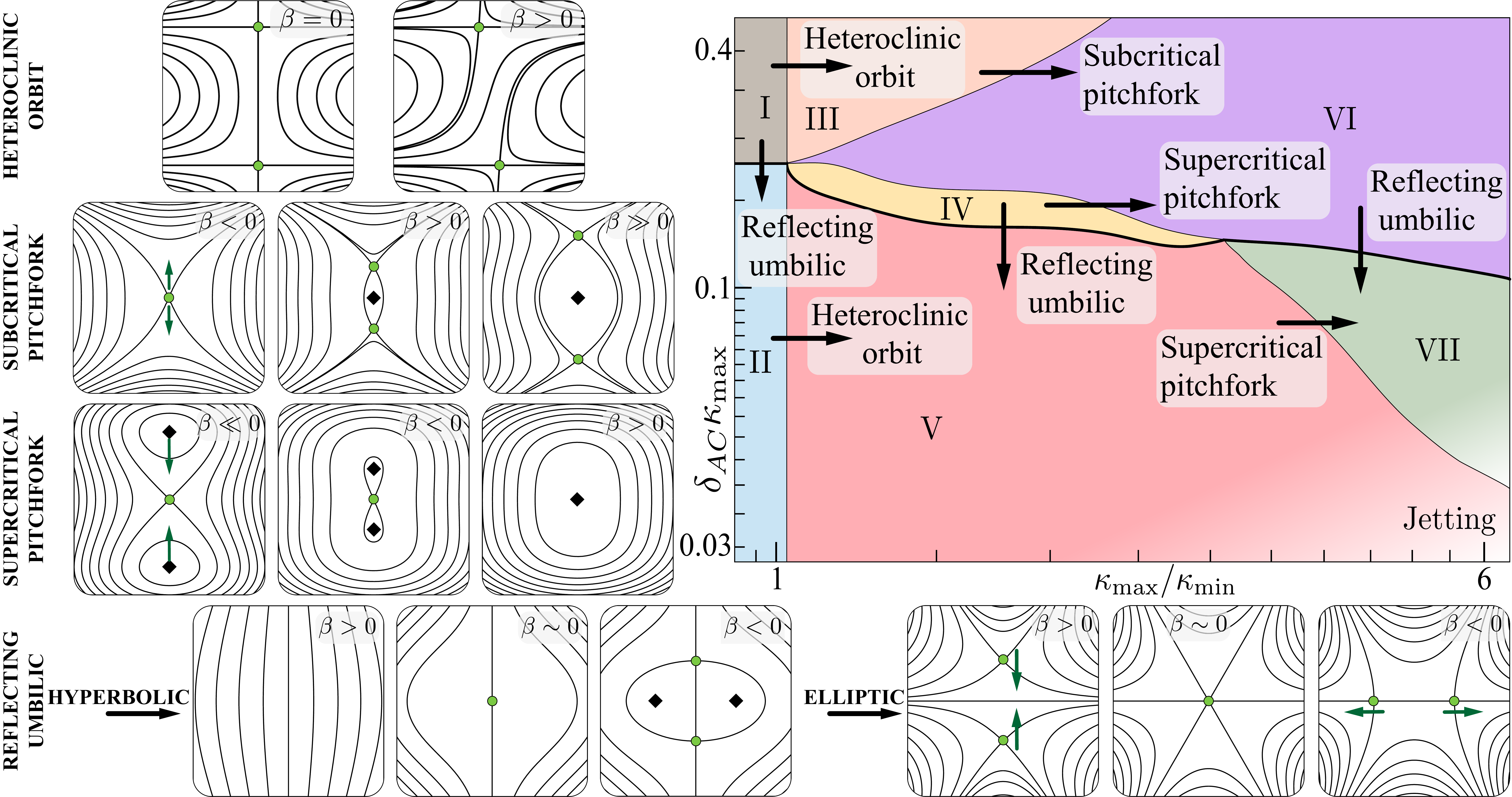}}
    \caption{Summary of the bifurcations seen in the lattice phase space.}\label{fig:bifsum}
\end{figure}

In the previous sections we identified all the bifurcations at play in our system, by
focusing on a few particular phase transitions. In figure \ref{fig:bifsum} we classify all
phase transitions, completing our analysis. 
Therefore, figure \ref{fig:bifsum} provides a compact
rulebook to manipulate streaming flows based on curvature
($\kappa_{\textrm{\scriptsize{max}}} / \kappa_{\textrm{\scriptsize{min}}}$) and background
oscillatory flow ($\delta_{AC} \kappa_{\textrm{\scriptsize{max}}}$) variations.

\vspace{-5pt}
\section{Generalization to individual convex streaming bodies}\label{sec:des}
\vspace{-5pt}

We now hypothesize that the above insights generalize to convex complex shapes immersed in an 
unbounded flow.
When considering a given shape, we proceed as follows: we identify a local
structure of interest, map it onto our phase space, predict how it will evolve based on local
body curvature change or background flow variation, and verify the outcome by comparing
with experiments and simulations.
\vspace{-10pt}
\subsection{Comparison against experiments: streaming triangles and squares}\label{sec:tat}
\begin{figure}[h!]
    \centerline{\includegraphics[width=\textwidth]{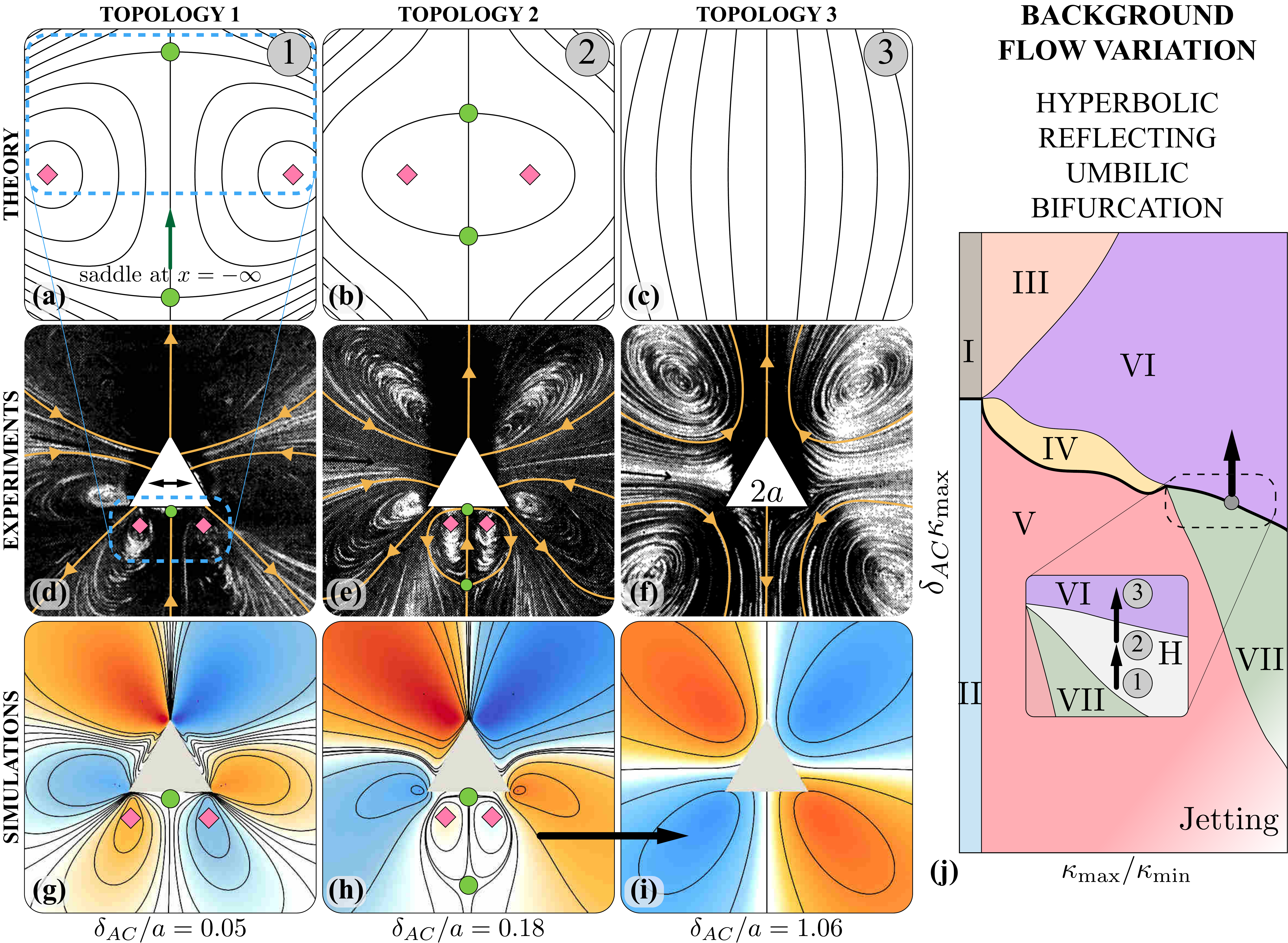}}
    \caption{Background flow variation for a triangle shaped cylinder.
        (\textit{a, b, c}) present the reduced Hamiltonian form contours for hyperbolic 
        reflecting umbilic bifurcation, associated with the transition hidden Phase H $\to$
        Phase \rom{6} in the lattice phase space. A topologically 
        equivalent transition is observed on varying $\delta_{AC} / a$, 
        both in experiments (\textit{d, e, f}), and simulations
        (\textit{g, h, i}).
        (\textit{j}) Mapping of the observed transition on the lattice phase space.}\label{fig:tat_tri}
\end{figure}

\vspace{-10pt}
We first consider an individual equilateral triangle (of side length $2a$), an object
characterized by top-down asymmetry and extreme ratio of curvatures, from sharp vertices
($\kappa_{\textrm{\scriptsize{max}}} \to \infty$) to flat sides
($\kappa_{\textrm{\scriptsize{min}}} \to 0$). In the original experiments performed by
Tatsuno (1975), this geometry was subject to oscillations and
three different flow topologies were observed for increasing $\delta_{AC} / a$, from 0.05
to 1.06 (figure \ref{fig:tat_tri}(\textit{d-f})).
In figure \ref{fig:tat_tri}(\textit{d}) we focus on the highlighted saddle and the two centres
near the horizontal edge of the triangle. This
structure closely resembles the hidden Phase H of figure \ref{fig:lat_umbcomb}(\textit{d, g}),
where the second saddle (not imaged in experiments) approaches from infinity 
(figure \ref{fig:tat_tri}(\textit{a})). 
We then map this structure onto our phase space (figure \ref{fig:tat_tri}(\textit{j})) and 
employ our previous analysis to predict the behavior of these critical points as
$\delta_{AC} / a$ is increased, similar to the experiments.
Based on our phase space (figure \ref{fig:bifsum}), as $\delta_{AC} / a$ increases the 
distance between critical points reduces,
bringing the saddle within the imaged domain (figure
\ref{fig:tat_tri}(\textit{b})) and forming a closed recirculating region, near the horizontal
edge. Upon a further increase of $\delta_{AC} / a$, we predict that the system will transition 
to a new topology corresponding to Phase \rom{6}, via a hyperbolic 
reflecting umbilic bifurcation (figure \ref{fig:tat_tri}(\textit{c})). This is a
consequence of the saddles and centres moving closer and closer, eventually collapsing and
vanishing.
Both experiments and simulations indicate that the system indeed behaves according to this
picture, confirming our predictions (figure \ref{fig:tat_tri}(\textit{d-i})).

\begin{figure}[h!]
    \centerline{\includegraphics[width=\textwidth]{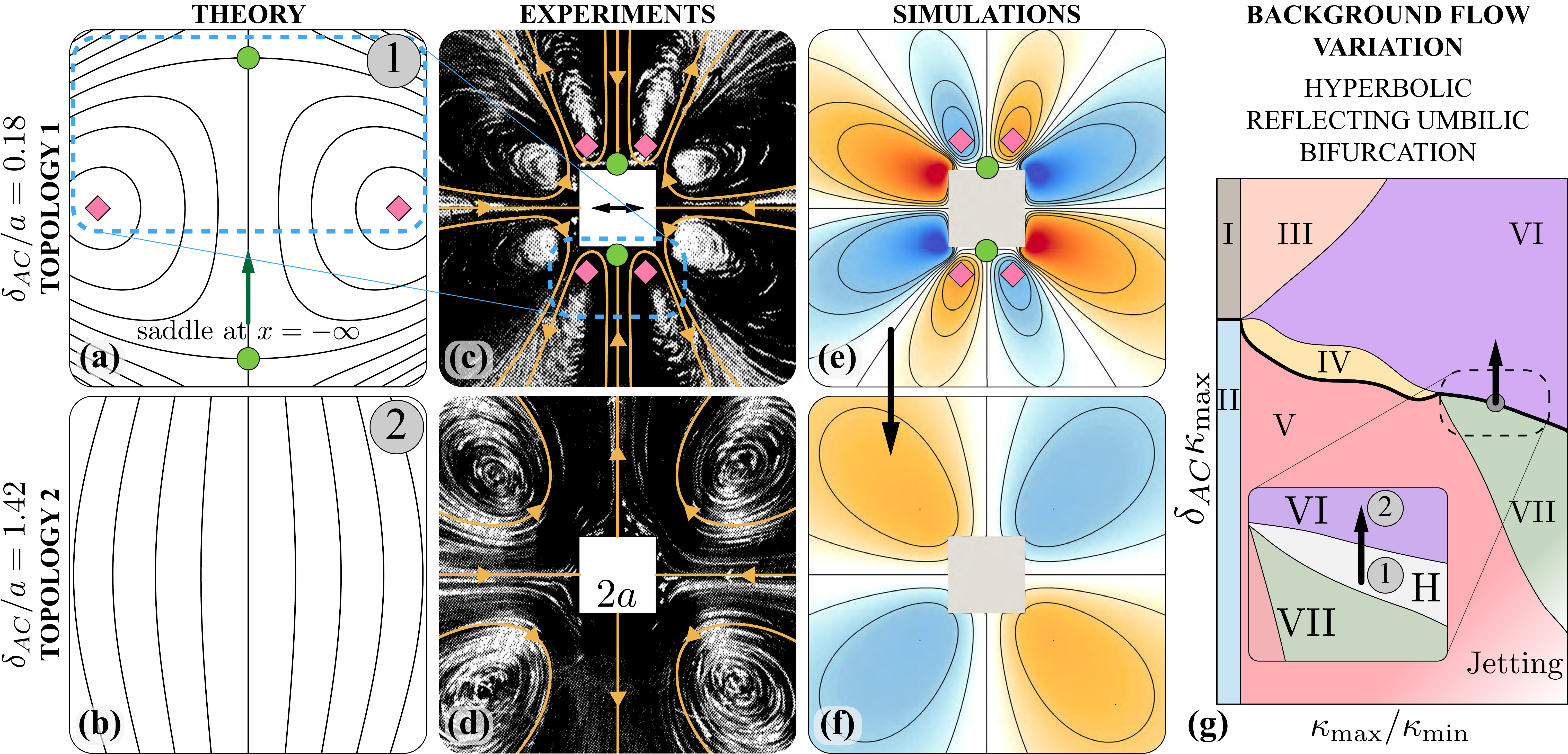}}
    \caption{Background flow variation for a square shaped cylinder.
        (\textit{a, b}) present the reduced Hamiltonian form contours for hyperbolic reflecting 
        umbilic bifurcation, associated with the transition hidden Phase H $\to$ Phase
        \rom{6}
        in the lattice phase space. A topologically 
        equivalent transition is observed on varying $\delta_{AC} / a$, 
        both in experiments (\textit{c, d}) and simulations
        (\textit{e, f}). 
        (\textit{g}) Mapping of the observed transition on the lattice phase space.}\label{fig:tat_sq2}
\end{figure}

\begin{figure}[h!]
    \centerline{\includegraphics[width=\textwidth]{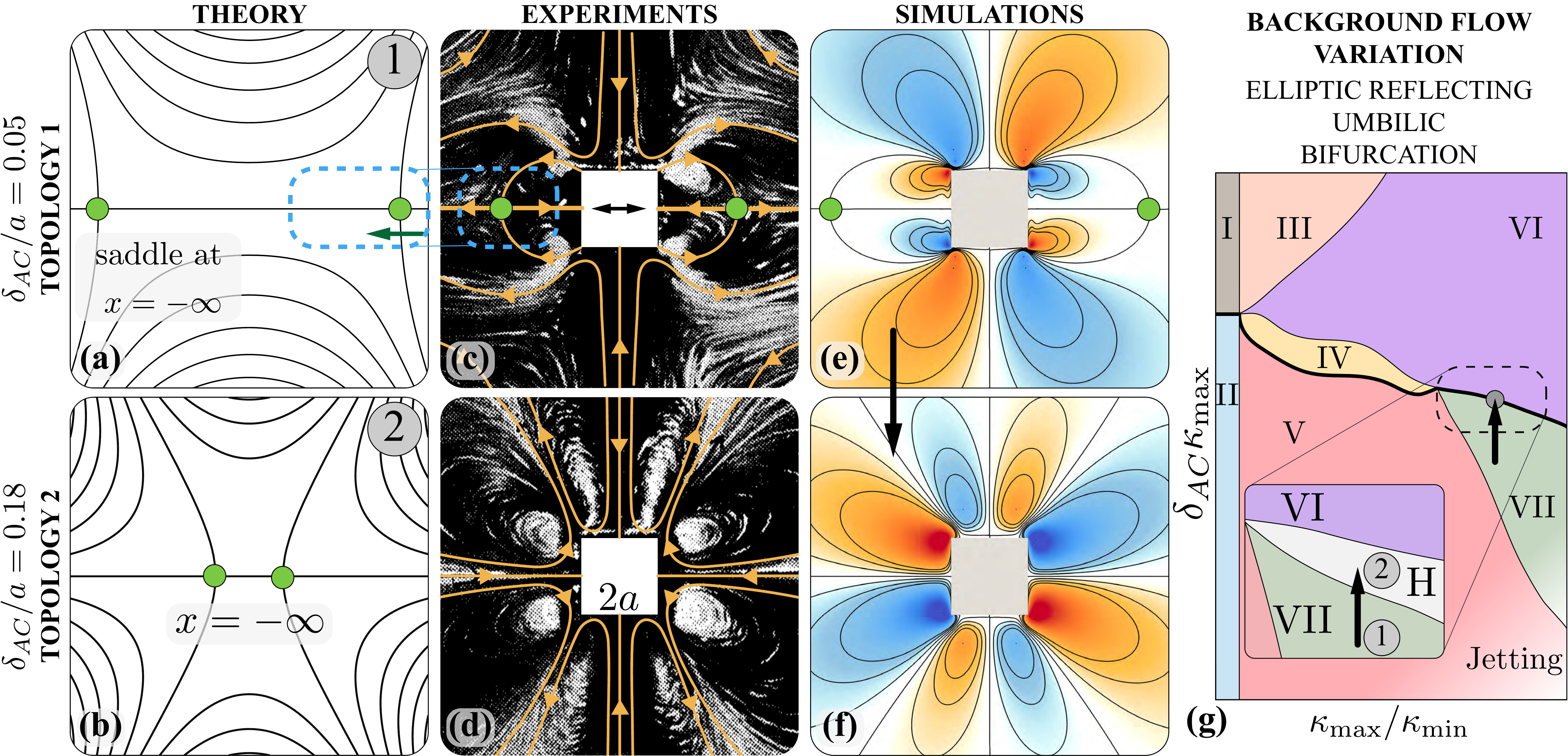}}
    \caption{Background flow variation for a square shaped cylinder.
        (\textit{a, b}) present the reduced Hamiltonian form contours for elliptic reflecting 
        umbilic bifurcation, associated with the transition \rom{7} $\to$
        hidden Phase H in the lattice phase space. A topologically 
        equivalent transition is observed on varying $\delta_{AC} / a$, 
        both in experiments (\textit{c, d}) and simulations (\textit{e, f}).
        (\textit{g}) Mapping of the observed transition on the lattice phase space.}\label{fig:tat_sq1}
\end{figure}

We further test our understanding, this time
against experiments involving an oscillating square cylinder (of side $2a$)
\citep{tatsuno1974circulatory}. 
Similar to the case of the triangle, three different flow topologies are observed for 
increasing $\delta_{AC} / a$. Figures \ref{fig:tat_sq2}(\textit{c, d}) and 
\ref{fig:tat_sq1}(\textit{c, d}) report the corresponding experimental recordings.
We first consider the case of figure \ref{fig:tat_sq2}(\textit{c, d}) in which
$\delta_{AC} / a$ was varied from 0.18 to 1.42. This case closely resembles the dynamics
associated with the triangle: indeed the highlighted saddles and centres (this time near
both the top and bottom horizontal edges) can be mapped to the hidden Phase H of 
figure \ref{fig:lat_umbcomb}(\textit{d, g}) and, as $\delta_{AC} / a$ increases, undergo the
same hyperbolic reflecting umbilic bifurcation, annihilating each other.
Again, predictions, experiments and simulations agree (figure \ref{fig:tat_sq2}(\textit{c-f})).
We note that the fact that identical local geometrical features affect the flow in a consistent
fashion across globally different shapes (triangles and squares) points at 
the robustness of our approach.
Finally, we consider the case of figure \ref{fig:tat_sq1}(\textit{c, d}) of a square cylinder 
at $\delta_{AC} / a = 0.05$ and 0.18.
We focus on the highlighted saddles forming a recirculation zone near the vertical sides
of the square. 
This structure closely resembles Phase \rom{7} of figure \ref{fig:lat_umbcomb}(\textit{k, n}),
where the second saddle (not imaged in experiments) is located at infinity.
Thus, as $\delta_{AC} / a$ increases, we predict that the 
saddles near the square will progressively move outwards, to approach the
saddles at infinity and undergo an elliptic reflecting umbilic bifurcation 
(figure \ref{fig:tat_sq1}(\textit{b})). This has the overall effect to enlarge the 
recirculation zone on the sides of the square. Once again, experiments and simulations
confirm our intuition (figure \ref{fig:tat_sq1}(\textit{c-f})). 

We note here that in all these cases Phase H is not as narrow as in our lattice system. 
This is not inconsistent: indeed we expect the phase boundaries to shift quantitatively
for globally different geometries, and our analysis holds as long as the phase space
structure is locally preserved. Although we cannot mathematically prove that the phase 
space organization is retained in any generic setting, we complemented the investigations 
presented here with a number of other 
studies (presented in the appendix for brevity), and all of them 
were found to be consistent with our analysis. This empirical validation 
underscores the practical use of our approach for flow design and manipulation purposes, 
as further exemplified in the next section.
\vspace{-10pt}
\subsection{Rational design of a streaming-enhanced transport bot}\label{sec:bullet}
\vspace{-10pt}
\begin{figure}[h!]
    \centerline{\includegraphics[width=\textwidth]{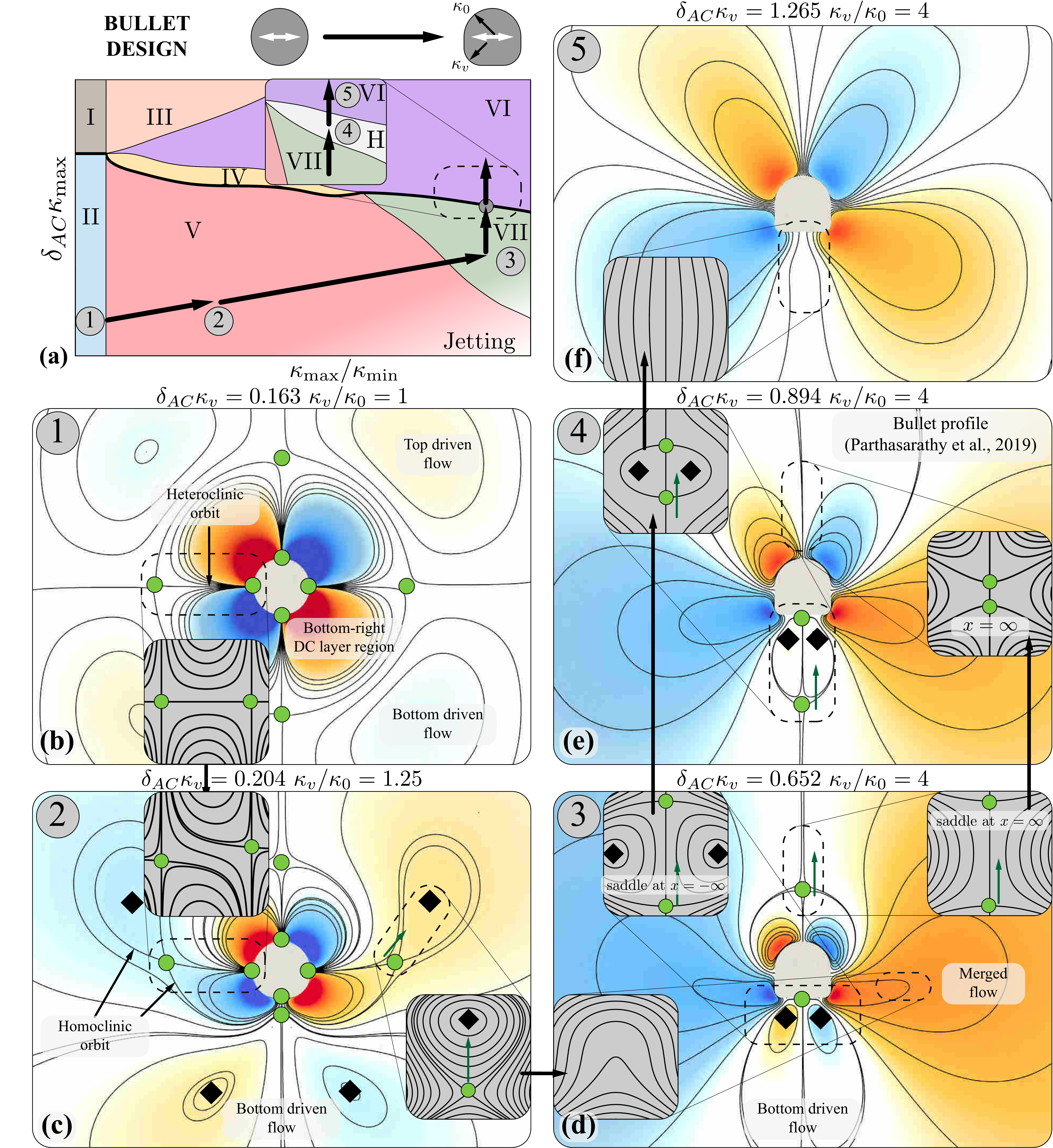}}
    \caption{Flow topology manipulation. (\textit{a}) Illustration of morphing a circular
        cylinder into a circular--square hybrid shaped cylinder with a circular side and a
        square side. Mapping of the observed transitions on the
        lattice phase space. (\textit{b-f}) Different topologies observed on
        geometric and background flow variation, with the concerned critical points
        highlighted and the predictions (reduced Hamiltonian form contours) illustrated as
        grey contours.}\label{fig:bullety_ps}
\end{figure}

In a recent study, we illustrated how a `bullet' shaped streaming bot enhances fluid
mediated transport of passive particles relative to simple circular cylinders
\citep{parthasarathy2019streaming}. The bullet geometry was empirically determined 
based on experiments on triangles \citep{tatsuno1975circulatory}, from which we 
borrowed fore-aft asymmetry and high rear curvature profiles. The rationale was to reproduce 
the closed streaming recirculation region observed in figure \ref{fig:tat_tri}(\textit{e,h}) 
to effectively trap trailing particles  and favor their transport as the bot displaces forward.
Here, we elucidate how that flow topology 
and object geometry could have been rationally designed based on our phase space, in a
step by step fashion.

Figure \ref{fig:bullety_ps}(\textit{a}) illustrates our design process of morphing
a circular cylinder into a circular--square hybrid cylinder (bullet). 
This hybrid cylinder presents top-down asymmetry---the top side is a circle with constant
curvature $\kappa_0$, while the bottom is a square with rounded corners of constant curvature $\kappa_v$. 
Hence specifying $\kappa_v / \kappa_0$ characterizes the shape geometry---
for $\kappa_v / \kappa_0 = 1$ the body is a perfect circle, while for $\kappa_v / \kappa_0 > 1$ 
the body presents a range of curvatures $\in [0, \kappa_v]$ on its bottom half. To
completely characterize this system, we capture 
the background flow variation using  $\delta_{AC} \kappa_v$, similar to the lattice phase space.

In the following, we break down the morphing design process in s.pdf. Each one of them relates 
a geometric or background flow variation to a corresponding local flow topology change, for which we highlight 
the concerned critical points and orbits. 

\vspace{-15pt}
\subsubsection{Step \circled{1} $\to$ \circled{2}}
\vspace{-15pt}

We start by choosing a $\delta_{AC} \kappa_{\scriptstyle \text{max}}$ for which the streaming flow
topology for a circular cylinder ($\kappa_v / \kappa_0 = 1$) lies in the finite-thickness 
DC layer regime. This corresponds to Phase \rom{2} in figure
\ref{fig:bullety_ps}(\textit{a}). With in mind the goal of reshaping the overall flow
topology to mimic the favourable features of figure \ref{fig:tat_tri}(\textit{e, h}), 
the first step is to ``open up" the DC layer. We then focus on the highlighted heteroclinic orbits in figure
\ref{fig:bullety_ps}(\textit{b}).
Recalling our observations in the lattice phase
space, we predict that breaking top-down symmetry ($\kappa_v / \kappa_0 > 1$) will break
these orbits via a heteroclinic orbit bifurcation (Phase \rom{2} $\to$ \rom{5} of figure
\ref{fig:bifsum}). This is computationally confirmed in figure
\ref{fig:bullety_ps}(\textit{c}), and has the effect of unfolding horizontally the two 
bottom recirculating regions of the DC layer.

\vspace{-15pt}
\subsubsection{Step \circled{2} $\to$ \circled{3}}
\vspace{-15pt}

Next, we need to prepare the conditions to form the closed flow region behind the cylinder
(similar to figure \ref{fig:tat_tri}(\textit{d, g})). This can be achieved by collapsing 
the newly generated saddles and centers. Their annihilation will cause the unfolded DC layer 
pockets to merge with the top driven flow regions. This, in turn, 
pushes the bottom driven flow regions against each other, compressing and aligning them
vertically (and eventually connecting them at infinity via a saddle point).
We then focus on the highlighted saddles and centres of figure
\ref{fig:bullety_ps}(\textit{c}), which closely resemble Phase \rom{5} of figure
\ref{fig:lat_suppitchf}(\textit{b}). We predict that a further increase in $\kappa_v /
\kappa_0$ will result in the saddle and centre approaching each other and eventually
colliding, leading to a saddle--centre bifurcation
(Phase \rom{5} $\to$ \rom{7} of figure \ref{fig:bifsum}). The occurrence of this
bifurcation, with the subsequent flow topology rearrangement, is indeed numerically
confirmed in figure \ref{fig:bullety_ps}(\textit{d}). 

\vspace{-15pt}
\subsubsection{Step \circled{3} $\to$ \circled{4}}
\vspace{-15pt}

Now the flow is favorably rearranged. In order to recover the useful closed 
flow region of figure \ref{fig:tat_tri}(\textit{e, h}), we need to bring closer to the
streaming body the saddle that connects at infinity the bottom driven flow regions.
We then focus on the highlighted saddles and two centres near the horizontal edge of the
bullet in figure \ref{fig:bullety_ps}(\textit{d}), which closely resembles the hidden
Phase H of figure \ref{fig:lat_umbcomb}(\textit{d}) (with one saddle at infinity) or
figure \ref{fig:tat_tri}(\textit{a, d}). Then a background flow variation ($\delta_{AC} \uparrow$)
pulls the saddle upwards, closer to the body, as predicted (figure
\ref{fig:bullety_ps}(\textit{e})). Additionally, we note that the flow structure in front 
(top) of the streaming shape can be mapped to Phase \rom{7} of figure
\ref{fig:lat_umbcomb}(\textit{k}). Therefore, as a side effect of the increase in
$\delta_{AC}$, the saddle in front of the body is pushed away upwards, eventually (next step) 
undergoing the elliptic reflecting bifurcation (Phase \rom{7} $\to$ hidden Phase H)
observed in figure \ref{fig:lat_umbcomb}(\textit{j}) and figure
\ref{fig:tat_sq1}(\textit{d, f}). Computations of figure \ref{fig:bullety_ps}(\textit{e}) 
confirm this intuition.

\vspace{-15pt}
\subsubsection{Step \circled{4} $\to$ \circled{5}}
\vspace{-15pt}

Finally, in addition to controlling the size of the closed flow region as discussed above, 
we now demonstrate (for completeness) how we can further manipulate its presence or
absence. 
We focus on the highlighted saddles and the two centres
near the horizontal edge of the bullet (figure \ref{fig:bullety_ps}(\textit{e})), which
closely resemble the hidden Phase H of figure \ref{fig:lat_umbcomb}(\textit{d}).
Then a background flow variation ($\delta_{AC} \uparrow$)
results in the saddles and centres approaching each other to then collapse, thus making
the closed flow region disappear. This is computationally confirmed in figure
\ref{fig:bullety_ps}(\textit{f}), which validates our prediction of a hyperbolic reflecting 
umbilic bifurcation (hidden phase H $\to$ Phase \rom{6}, figure
\ref{fig:lat_umbcomb}(\textit{b}), figure \ref{fig:tat_tri}(\textit{f, i}) and
figure \ref{fig:tat_sq2}(\textit{d, f})).

\vspace{3mm}

In summary, this section illustrates how our approach can be employed to
predict and design streaming flow topologies, in a rational fashion.

\vspace{-5pt}
\section{Conclusions}\label{sec:con}
\vspace{-5pt}

With the goal of extending our understanding of streaming phenomena to include body curvature 
effects, we propose a simplified setting in which multiple circular cylinders are regularly 
arranged in a periodic lattice. We systematically investigate this system to construct a 
phase space that relates local body curvature and background flow variations to streaming 
flow topology. The obtained phase space reveals rich dynamics on account of the
non-linear, collective behavior that stems from the presence of multiple body length
scales. The phase space is subsequently analyzed through the lens of 
dynamical system theory, to detect the bifurcations and physical mechanisms at play. 
We then demonstrate that our understanding in the simplified lattice system can be extended to 
individual bodies presenting a spectrum of curvatures. Altogether these results provide physical 
intuition and a rulebook to manipulate and design streaming flow topologies, which may find 
useful application in microfluidics and micro-robotics.

Although our study provides a systematic prediction of flow topology transitions on geometric 
and background flow variation, our understanding still remains incomplete. 
How do we incorporate concave geometric features?
How does our approach extend to three-dimensional settings? 
How does body topology affect flow topology? These questions are 
beyond the scope of the current paper, and are avenues of future research.

\vspace{-5pt}
\section{Acknowledgements}\label{sec:ack}
\vspace{-5pt}

We thank Sascha Hilgenfeldt for helpful discussions over the course of this work.
The authors acknowledge support by the National Science Foundation under NSF
CAREER Grant No. CBET-1846752 (MG) and by the Blue Waters project (OCI-
0725070, ACI-1238993), a joint effort of the University of Illinois at Urbana-Champaign
and its National Center for Supercomputing Applications. 
This work used the Extreme Science and Engineering Discovery Environment (XSEDE)
\citep{towns2014xsede} Stampede2, supported by National Science Foundation grant 
number ACI-1548562, at the Texas Advanced Computing Center (TACC) through allocation TG-MCB190004.

\vspace{-5pt}
\section{Appendix}
\vspace{-5pt}
\appendix

\vspace{-5pt}
\section{Curvature variation setup: cylinders in an infinite regular lattice}
\vspace{-5pt}

Here we provide further details relative to our setup, which consists of a
lattice of cylinders with two discrete curvatures. With a computational domain of
range $[0, 1]$, we instantiate a lattice of 16 cylinders in a $4 \times 4$ grid with periodic
boundary conditions. To accommodate 4 cylinders in our domain, while allowing a curvature 
variation up to 
$\kappa_{\textrm{\scriptsize{max}}}/\kappa_{\textrm{\scriptsize{min}}} = 6$, 
we fix $\kappa_{\textrm{max}} = 0.02$ as a reference length scale. Then 
the constant centre-to-centre distance
between these cylinders is $s = 1 / 4 = 0.25 = 12.5 \kappa_{\textrm{max}}$.
Maintaining $s$ constant allows us to consistently compare systems characterized by different 
curvature ratios. The value of $s$ is chosen to span a reasonable curvature ratio range, and at 
the same time is practical from an experimental, fabrication standpoint. 
The oscillatory amplitude ($A$) for all the cylinders in the lattice is kept constant 
($A \kappa_{\textrm{\scriptsize{max}}} = 0.1$). 
This is because oscillating the fluid with
constant amplitude is experimentally convenient. By virtue of keeping a constant amplitude, 
the streaming flow for the 
lattice system can be characterized with one value of
$R_s$ ($R_s = A^2 \omega / \nu$). 
Lastly, fixing a constant amplitude allows us to extend our
understanding (with two curvatures) to individual complex shapes with multiple
curvatures. Indeed all local curvatures of an individual streaming body undergo the same
absolute oscillation amplitudes. We not here that we performed cursory phase space 
explorations for different spacings between cylinders, and observed that the qualitative 
nature of the emerging streaming fields is preserved, although the boundaries between 
different topological phases shift quantitatively. 

\vspace{-5pt}
\section{Lattice system: flow bifurcations}
\vspace{-5pt}

\subsection{Heteroclinic orbit bifurcation via asymmetric background flow}
\vspace{-10pt}

\begin{figure}[h!]
    \centerline{\includegraphics[width=\textwidth]{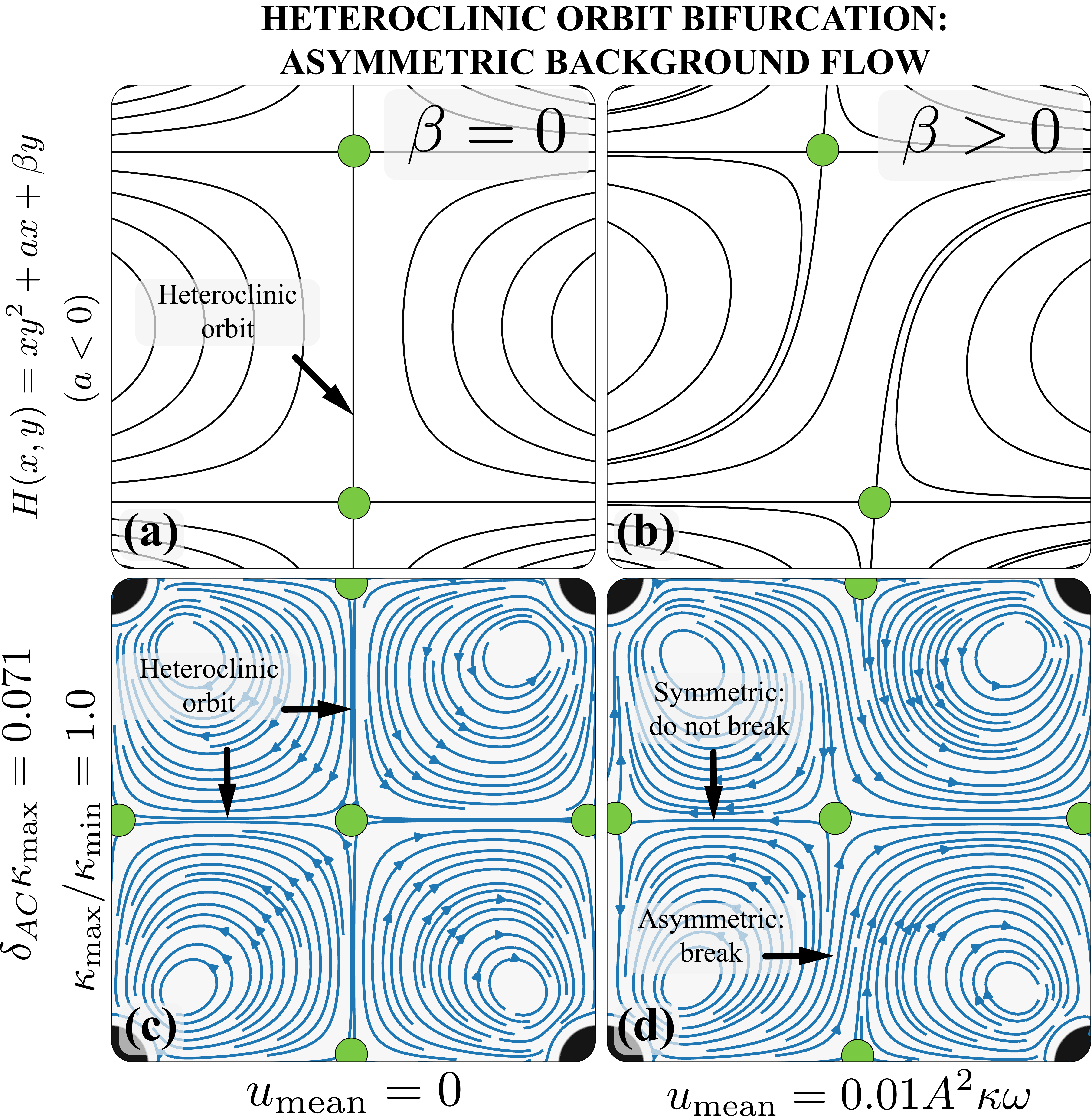}}
    \caption{Asymmetric background flow. We demonstrate that the heteroclinic orbit
    bifurcation, associated with the transition \rom{2} $\to$
   	\rom{5} in the lattice phase space, can be triggered by breaking symmetry via
  	superimposing a slow uniform background flow, instead of using curvature
   	variations as in the main text. (\textit{a, b}) present the reduced Hamiltonian form
    for the bifurcation, with the flow topology change on imposing the mean flow in
	the lattice system, shown in (\textit{c, d}). We note that in (\textit{c, d}) all cylinders present
	the same radius so that $\kappa_{\textrm{max}} / \kappa_{\textrm{min}} = 1$.}
\label{fig:lat_hetero_mf}
\end{figure}

Here we revisit the heteroclinic orbit bifurcation observed for the transition
\rom{2} $\to$ \rom{5} of figure 4 in the main text. We identified the reduced Hamiltonian form with 
the unfolding term coefficient $\beta$ that represents symmetry ($\beta = 0$) or
asymmetry ($\beta \neq 0$) in our lattice setup 
(figure \ref{fig:lat_hetero_mf}(\textit{a, b})). We now demonstrate that the 
notion of breaking symmetry extends to more than just curvature variation, by breaking 
symmetry via the superposition of a uniform background flow. Figure \ref{fig:lat_hetero_mf}(\textit{c})
presents a Phase \rom{2} flow topology. We focus on the highlighted saddles
and the heteroclinic orbits joining them. As we break symmetry along the horizontal axis
by imposing a slow mean flow, we observe the predicted flow topology change 
(figure \ref{fig:lat_hetero_mf}(\textit{d})), where the 
vertically oriented heteroclinic orbits break up, while the horizontally oriented ones do not.   
Through this illustration, we demonstrate that the control on asymmetry
via means other than curvature variation leads to identical flow topology
rearrangements. This confirms the physical intuiton provided by the identified reduced
Hamiltonian form.  

\vspace{-10pt}
\subsection{Reflecting umbilic bifurcation: two step rationale}
\vspace{-10pt}

\begin{figure}[h!]
    \centerline{\includegraphics[width=\textwidth]{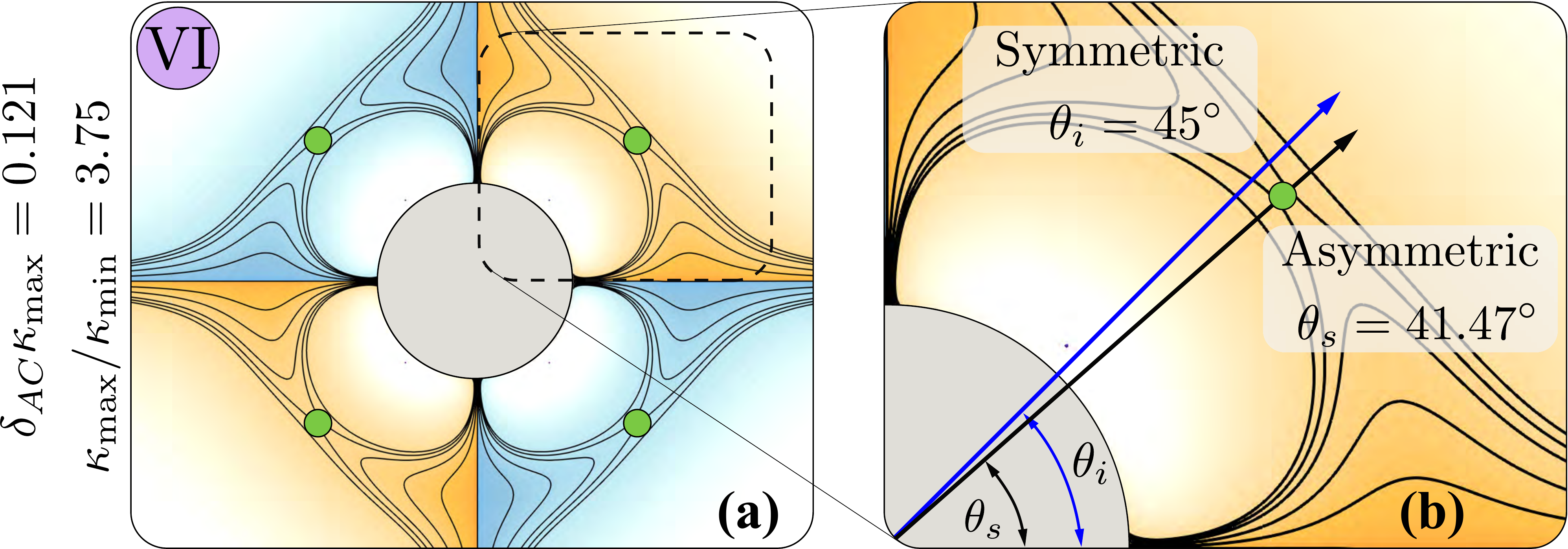}}
    \caption{Reflecting umbilic bifurcation. (\textit{a}) Illustration of the flow topology
    near the smaller cylinder for Phase \rom{6}. (\textit{b}) Angular location of the highlighted
saddle observed in simulations $\theta_s$, compared to the expected symmetric location
$\theta_i$. This small yet non-negligible difference creates asymmetry between the flow
topology along the vertical and horizontal axes, resulting in the two step reflecting
umbilic bifurcation. }\label{fig:umb_rat}
\end{figure}

We now focus on the transition between Phase \rom{6}
and \rom{7} in the lattice phase space. We notice that the system appears symmetric about the horizontal and 
vertical axes. Yet, the transition into Phase \rom{7} occurs in two distinct s.pdf: 1) a
hyperbolic reflecting umbilic bifurcation that involves a topological rearrangement
about the vertical axis of symmetry of the cylinder, and 2) an elliptic reflecting
umbilic bifurcation that involves a topological rearrangement about the horizontal axis
of symmetry of the cylinder. To 
investigate this preference for one axis over another, we focus on 
figure \ref{fig:umb_rat}(\textit{a}), which depicts the flow topology near the
smaller cylinder in Phase \rom{6}. We focus on the highlighted saddles and the
corresponding local flow topology depicted in figure \ref{fig:umb_rat}(\textit{b}). We
notice a small yet non-negligible difference between the angular location of the highlighted
saddle in simulations ($\theta_s$), and the expected location ($\theta_i$) 
based on the symmetry of the setup. Upon increasing simulation's resolution, or varying
background flow oscillation direction, the magnitude of this angular deviation can be
reduced, but not removed, so that a small preference persists. We attribute this small
preferential deviation to the fact that the cylinders location is not (and cannot be) perfectly aligned
with the discretization grid.

This small local deviation from symmetry is responsible for the observed sequence of
umbilic bifurcations. The sequence order is found to be robust: indeed a saddle
misalignment with respect to the $45^{\circ}$ angle consistently triggers the hyperbolic
reflecting umbilic bifurcation first (as $\delta_{AC} \kappa$ decreases), either on the 
vertical or horizontal axis, depending on $\theta_s < 45^{\circ}$ or $\theta_s > 45^{\circ}$. 
The induced local topological rearrangement is then found to be the natural precursor of 
the elliptic reflecting umbilic bifurcation, which occurs on the axis perpendicular to 
the first bifurcation. Therefore, to summarize, the boundaries between Phase \rom{6}, 
hidden Phase H and Phase \rom{7} are robustly associated with the
bifurcation identified in the main text, as empirically verified though the comparison
with triangle and square experiments.
We note that in a perfect lattice, this series of two s.pdf may coalesce in a single step.
Nonetheless, this is only a theoretical scenario which we could not attain computationally
and that cannot be expected to manifest experimentally.

\vspace{-10pt}
\subsection{Phase \rom{1} $\to$ \rom{2}: higher order reflecting umbilic
bifurcation}\label{sec:symumb}
\vspace{-10pt}

\begin{figure}[h!]
    \centerline{\includegraphics[width=0.8\textwidth]{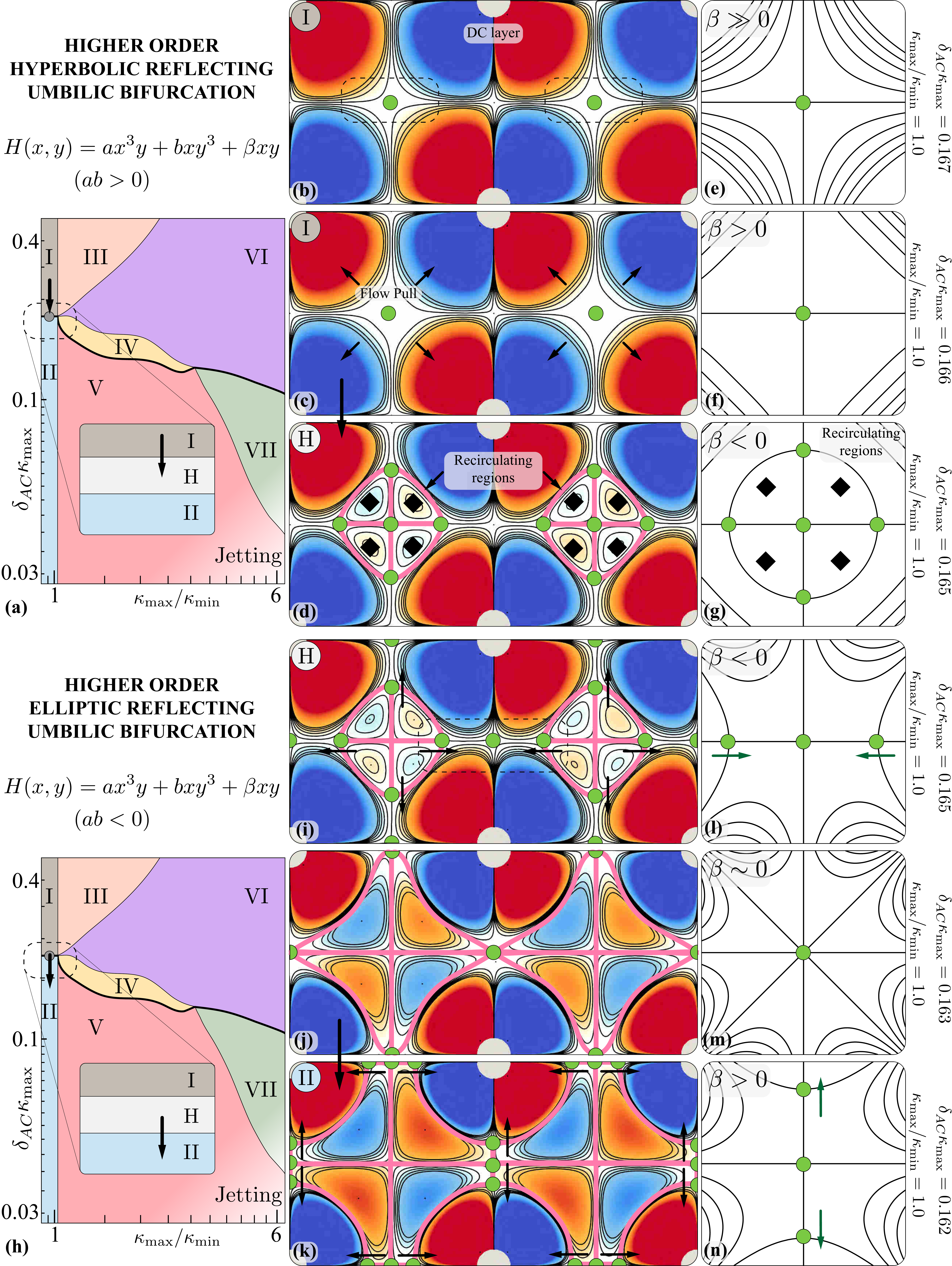}}
    \caption{
        Phase \rom{1} $\to$ hidden Phase H: higher order hyperbolic reflecting umbilic bifurcation. (\textit{a}) 
        The transition is highlighted on the phase space (with a zoomed in view) and the corresponding reduced
        Hamiltonian form is reported. (\textit{b, c, d}) Flows representative of Phase
        \rom{1}, Phase \rom{1} approaching the transition, and hidden Phase H, respectively. 
        (\textit{e, f, g}) Bifurcations captured as contours of the reduced Hamiltonian form.        
        (\textit{h}) The transition from hidden Phase H $\to$ \rom{2} (higher order elliptic reflecting
        umbilic bifurcation) is highlighted on the phase space (with a zoomed in view) and the 
        corresponding reduced Hamiltonian form is reported. (\textit{i, j, k}) Flows representative of
        hidden Phase H, at the transition, and Phase \rom{2}, respectively. 
        (\textit{l, m, n}) Bifurcations captured as
        contours of the reduced Hamiltonian form. The newly created recirculating region pairs are marked in
        pink.}\label{fig:lat_symumbcomb}
\end{figure}

Here, we illustrate the bifurcation Phase \rom{1} $\to$ \rom{2}, as depicted 
in figure \ref{fig:lat_symumbcomb}(\textit{a}). 
To identify this bifurcation we focus on two adjacent unit cell quadrants of the lattice. 
We note the absence of driven flow regions around the cylinders
in Phase \rom{1} (figure \ref{fig:lat_symumbcomb}(\textit{b})) and their presence in phase
\rom{2} (figure \ref{fig:lat_symumbcomb}(\textit{k}), marked in pink).
This flow topology change occurs in two consecutive s.pdf, passing through another hidden
phase of \S 4.2.8 in the main text. 

In the first step, we draw attention to the absence of recirculating regions
in Phase \rom{1} (figure \ref{fig:lat_symumbcomb}(\textit{b})) and their presence in
figure \ref{fig:lat_symumbcomb}(\textit{d}) (comprised of five saddles and four centres,
marked in pink). 
We note that the latter flow field corresponds to the hidden Phase H.  
The simplest Hamiltonian form that captures this transition is $H(x, y) = a x^{3} y + b x
y^{3} + \beta x y$ with $a b > 0$, which corresponds to a higher order hyperbolic
reflecting umbilic bifurcation (higher order with respect to the ones shown in the main
text, on account of the additional symmetry due to same curvature cylinders) 
\citep{bosschaert2013bifurcations}. Here $\beta x y$ is the unfolding term, that controls 
the appearance (going from $\beta > 0$ to $\beta < 0$) of the recirculating regions
and their size (figure \ref{fig:lat_symumbcomb}(\textit{e-g})).  
In our lattice system, the appearance and size of these regions can be
controlled by decreasing $\delta_{AC}$, which decreases the DC layer thickness
$\delta_{DC}$ of all the cylinders.
This pulls the streamlines adjacent to the marked saddle,
in four opposite directions (figure \ref{fig:lat_symumbcomb}(\textit{c})),
causing it to eventually split into five saddles and four centres
(figure \ref{fig:lat_symumbcomb}(\textit{d})). Topologically, this manifests as four 
counter-rotating recirculating regions, thus revealing Phase H.

The second step of the Phase \rom{1} $\to$ \rom{2} transition occurs upon further
decreasing $\delta_{AC}$, immediately after the previous bifurcation, thus 
rendering the hidden Phase H very narrow.
We focus on the saddles in the highlighted region of the unit cell, in the hidden Phase H (figure
\ref{fig:lat_symumbcomb}(\textit{i})).
After the transition these saddles are located on the vertical axis passing through the 
highlighted region, thus recovering  
Phase \rom{7} (figure \ref{fig:lat_symumbcomb}(\textit{k})).
The simplest Hamiltonian form that captures this rearrangement is $H(x, y) = a x^{3} y + b
x y^{3} + \beta x y$ with $a b < 0$, which corresponds to a higher order elliptic reflecting umbilic bifurcation 
\citep{bosschaert2013bifurcations}. Here $\beta xy$ is the unfolding term, that
captures whether the saddles are present ($\beta < 0$) or absent ($\beta > 0$) on the
horizontal axis (figure \ref{fig:lat_symumbcomb}(\textit{l-n})), as well as their distance.  
Similar to the previous step, a decrease in $\delta_{AC}$ (i.e $\delta_{DC}\downarrow$) causes a pull on the streamlines
immediately adjacent to the cylinder's DC layers. This time though, the saddles created
in the previous step are now pushed towards the saddle on the vertical axis in the
highlighted region, extending the recirculating region pairs. Upon reaching the
saddle on the vertical axis, the two opposite saddles collapse (figure \ref{fig:lat_symumbcomb}(\textit{j})) and
split along the vertical axis (figure \ref{fig:lat_symumbcomb}(\textit{k})). 
We note that this bifurcation is observed along all the edges of the unit cell.
These new saddles completely define the driven flow regions around the cylinders, thus
recovering Phase \rom{2}.

\vspace{-5pt}
\section{Generalization to individual streaming bodies: additional examples}
\vspace{-5pt}

Here we provide a number of additional examples which demonstrate that our intuition from the phase space 
carries on to various streaming geometries and oscillatory background flow conditions. 
All subsequent studies entail a single shape immersed in an unbounded domain.

\vspace{-10pt}
\subsection{Comparison against experiments: streaming triangles and diamonds}
\vspace{-10pt}

Expanding on the examples of the main text, we further 
test our understanding against a different set of experiments. We consider an oscillating diamond 
cylinder (of side $2a$) \citep{tatsuno1974circulatory} and a horizontally oriented
triangle cylinder (of side $2a$) \citep{tatsuno1975circulatory}. 

\begin{figure}[h!]
    \centerline{\includegraphics[width=\textwidth]{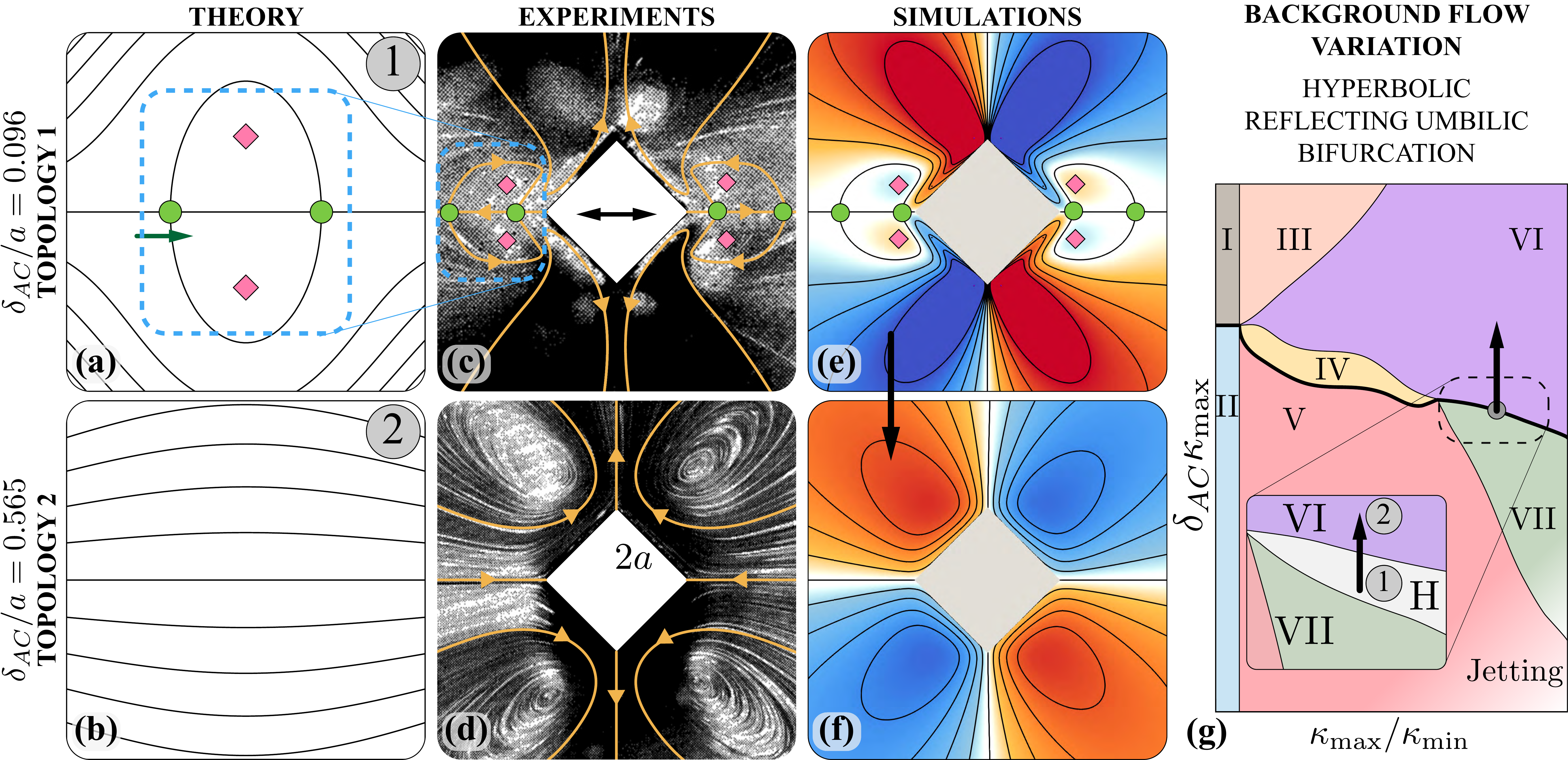}}
    \caption{Background flow variation for a diamond cylinder.
        (\textit{a, b}) Reduced
        Hamiltonian form contours for hyperbolic reflecting umbilic bifurcation, 
        associated with the transition hidden Phase H $\to$
        Phase \rom{6} in the lattice phase space. A topologically 
        equivalent transition is observed on increasing $\delta_{AC} / a$, 
        both in experiments (\textit{c, d}) and simulations
        (\textit{e, f}).
        (\textit{g}) Mapping of the observed transition on the lattice phase space.
}\label{fig:tat_dia}
\end{figure}

\begin{figure}[h!]
    \centerline{\includegraphics[width=\textwidth]{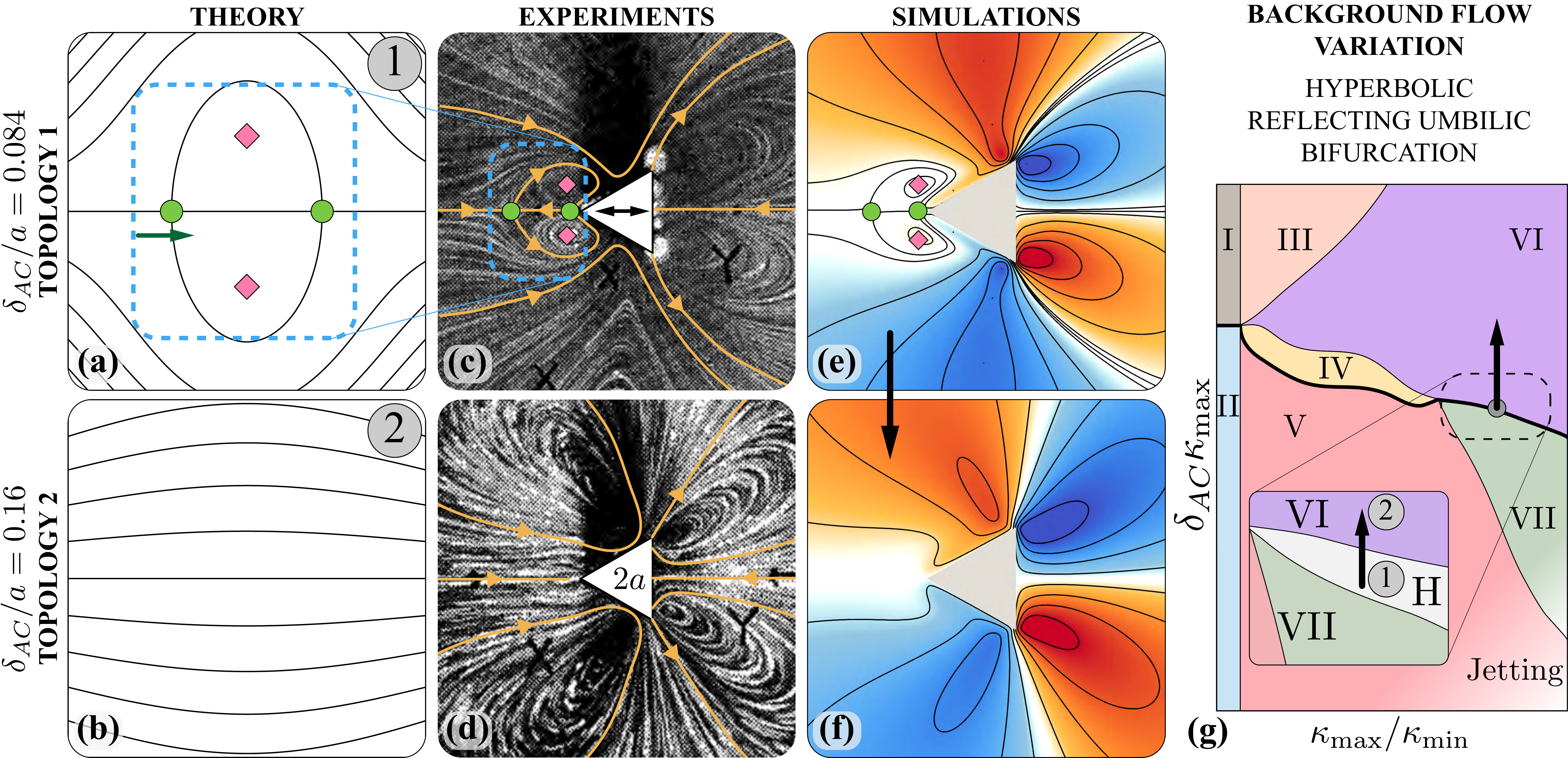}}
    \caption{Background flow variation for a horizontally oriented equilateral triangle 
        cylinder. (\textit{a, b}) Reduced
        Hamiltonian form contours for hyperbolic reflecting umbilic bifurcation, 
        associated with the transition hidden Phase H $\to$
        Phase \rom{6} in the lattice phase space. A topologically 
        equivalent transition is observed on increasing $\delta_{AC} / a$, 
        both in experiments (\textit{c, d}) and simulations (\textit{e, f}). 
        (\textit{g}) Mapping of the observed transition on the lattice phase space.
    }\label{fig:tat_htri}
\end{figure}

Figure \ref{fig:tat_dia} illustrates the hyperbolic reflecting umbilic bifurcation for a
diamond cylinder, observed on increasing $\delta_{AC} / a$ from 0.096 to 0.565, similar 
to the one observed in the lattice phase space.
Figure \ref{fig:tat_dia}(\textit{a, b}) illustrates the predicted flow topology
change based on the reduced Hamiltonian contours, with confirmations against flow
topologies observed in experiments (figure \ref{fig:tat_dia}(\textit{c, d}))
and simulations (figure \ref{fig:tat_dia}(\textit{e, f})).

Figure \ref{fig:tat_htri} illustrates the hyperbolic reflecting umbilic bifurcation for
the triangle cylinder, observed on increasing $\delta_{AC} / a$ from 0.084 to 0.16, 
similar to the one observed in the lattice phase space.
Once again experiments (figure \ref{fig:tat_htri}(\textit{c, d}))
and simulations (figure \ref{fig:tat_htri}(\textit{e, f})) confirm our
predictions.

\vspace{-10pt}
\subsection{Shape parametrization}\label{param}
\vspace{-10pt}

\begin{figure}[h!]
    \centerline{\includegraphics[width=\textwidth]{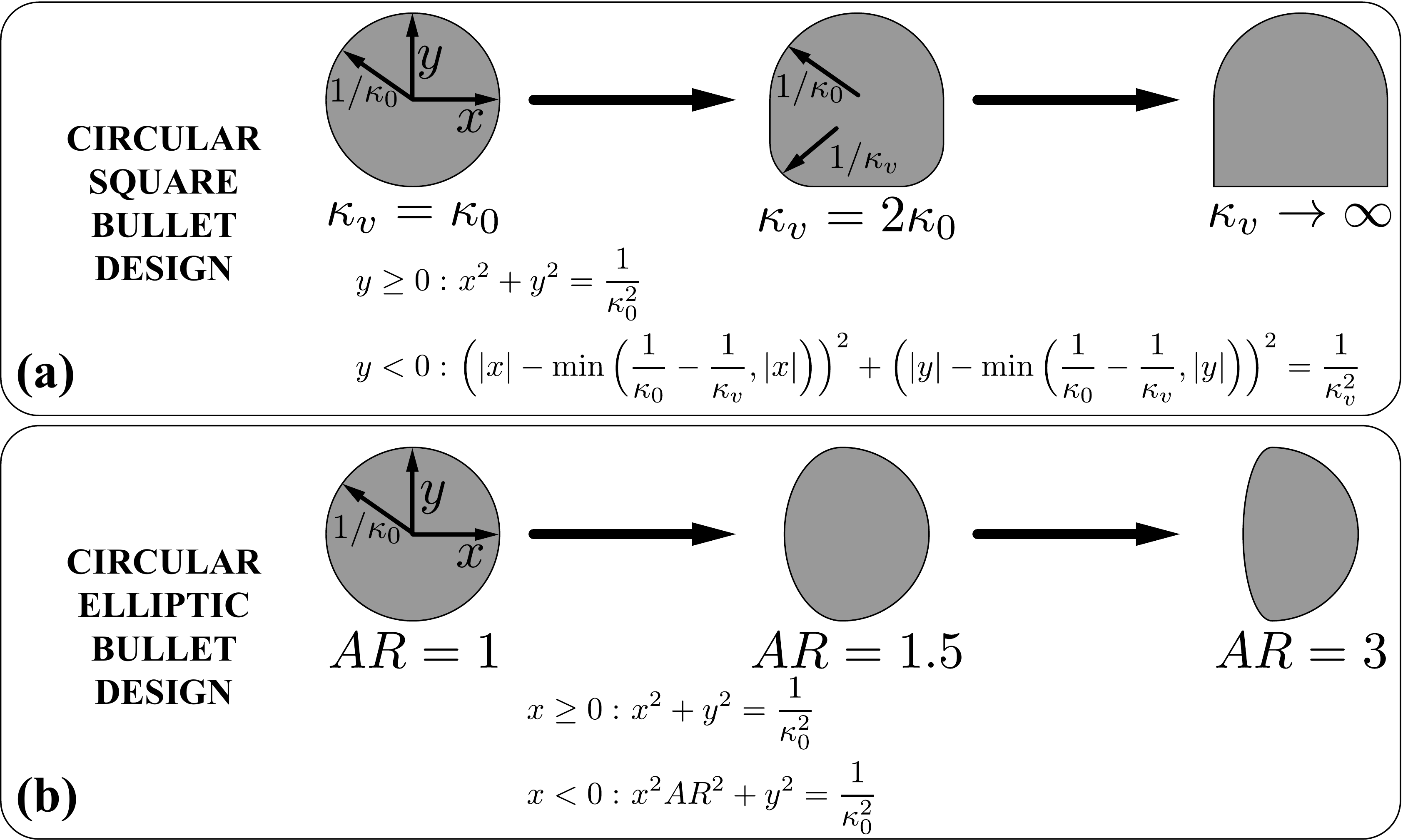}}
    \caption{ Parametrization of (\textit{a}) the hybrid circular--square cylinder (main text) and
        (\textit{b}) the circular--elliptic cylinder.
    }\label{fig:morph}
\end{figure}

Figure \ref{fig:morph} illustrates the parameterization for the complex convex shapes,
investigated in \S 5.2 of the main text and in \S \ref{vell_ps} of the appendix.  
Figure \ref{fig:morph}(\textit{a}) presents the parametrization for the hybrid
circular--square bullet of figure 12(\textit{a}) in the main text.
This hybrid cylinder presents top-down asymmetry---the top side is a circle with constant
curvature $\kappa_0$, while the bottom is a square with rounded corners of constant curvature $\kappa_v$. 
As $\kappa_v$ is varied from $\kappa_0$ to higher values ($\to \infty$), the shape morphs
from a circle to a circle--square hybrid with increasingly sharper corners.
Figure \ref{fig:morph}(\textit{b}) presents the parametrization for the hybrid
circular--elliptic cylinder, which we use to demonstrate additional examples of flow
topology design in the following sections.
This hybrid cylinder presents left-right asymmetry---
the right side is a circle with constant curvature $\kappa_0$, while the left side is an
ellipse of aspect ratio AR (defined as height/breadth, where height is kept constant). 
As AR is varied from 1 to higher values, the shape morphs
from a circle to a circular--elliptic cylinder with a higher AR elliptic side (with
corresponding higher curvature at the corners).

\vspace{-10pt}
\subsection{Manipulation of streaming flow topology: circle to a
circle--ellipse}\label{vell_ps}
\vspace{-10pt}

\begin{figure}[h!]
    \centerline{\includegraphics[width=\textwidth]{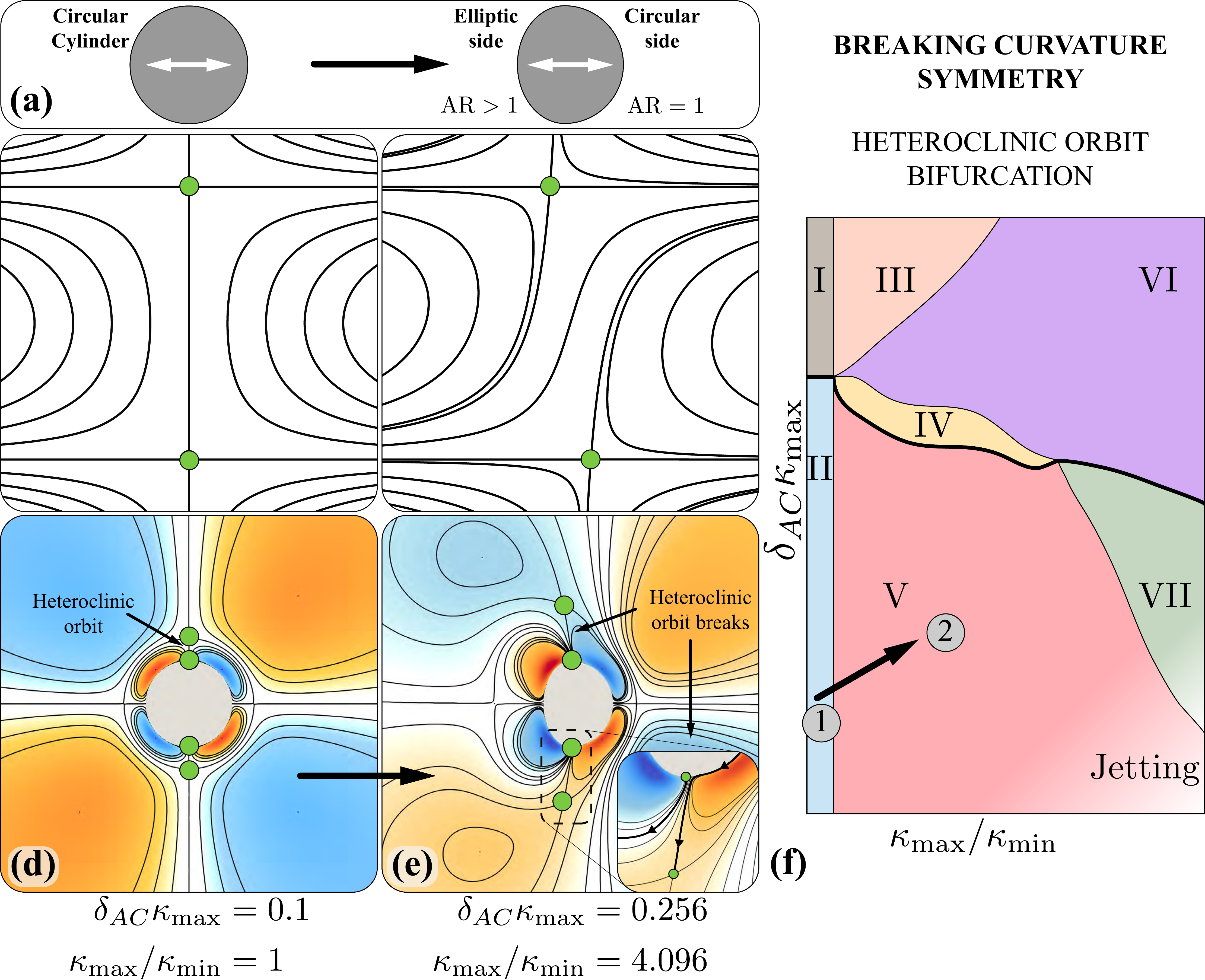}}
    \caption{Breaking symmetry. (\textit{a}) Illustration of morphing a circular
        cylinder into a hybrid circular--elliptic cylinder with a circular side and an elliptic
        side with varying aspect ratio ($\textrm{AR}$). (\textit{b, c}) present the reduced
        Hamiltonian form contours for heteroclinic orbit bifurcation, associated with 
        the transition \rom{2} $\to$ \rom{5} in the lattice phase space.
        The morphed body presents a topologically equivalent transition on breaking symmetry, presented in
        (\textit{d, e}). (\textit{f}) Mapping of the observed transition on the
        lattice phase space.
    }\label{fig:vell_hetero}
\end{figure}

Figure \ref{fig:vell_hetero}(\textit{a}) illustrates the morphing from a circular
cylinder of uniform aspect ratio ($\textrm{AR} = 1$, defined as height/breadth) to a body with a 
circular side having the same
curvature $\kappa_0$ and an elliptic side with aspect ratio $\textrm{AR} > 1$. 
Parametrization of this hybrid shape can be found in figure \ref{fig:morph}(\textit{b}). For
such $\textrm{AR}$, the body presents constant curvature $\kappa_0$ on the circular side
and a range of curvatures ($\kappa_0 / \textrm{AR}$ to $\textrm{AR}^2 \kappa_0$) on the
elliptic side, thus introducing curvature variation in a regulated fashion. 

\begin{figure}[h!]
    \centerline{\includegraphics[width=\textwidth]{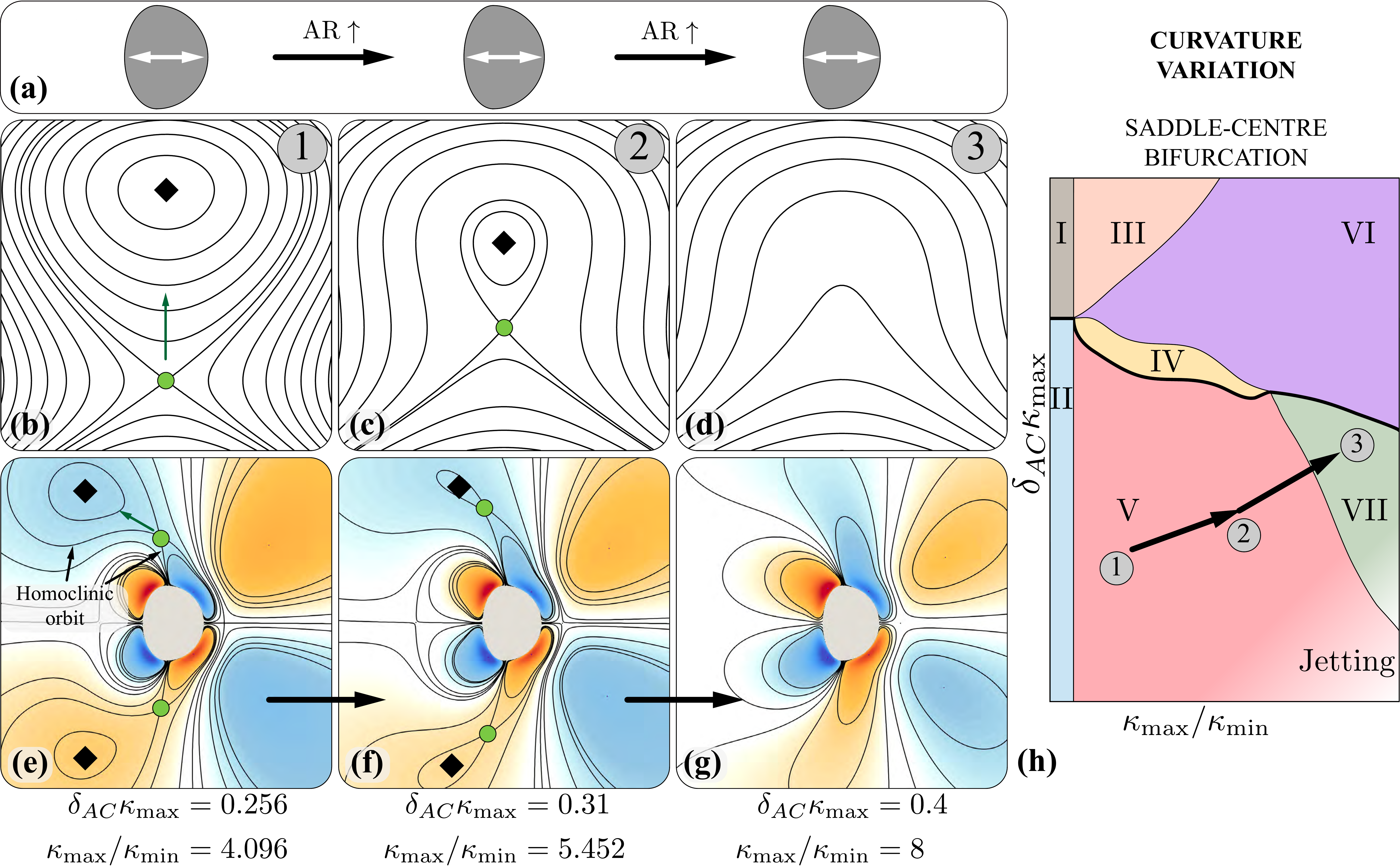}}
    \caption{Varying curvature ratio. (\textit{a}) Illustration of increasing curvature
        variation (increasing $\textrm{AR}$) for the circular--elliptic hybrid.
        (\textit{b, c, d}) present the reduced
        Hamiltonian form contours for saddle--centre bifurcation, similar to the supercritical
        pitchfork bifurcation, associated with the transition \rom{5} $\to$
        \rom{7} in the lattice phase space. A topologically 
        equivalent transition is observed
        on morphing the body, presented in (\textit{e, f, g}). 
        (\textit{h}) Mapping of the observed transition on the
        lattice phase space.
    }\label{fig:vell_sadcent}
\end{figure}

We start by choosing a $\delta_{AC} \kappa_{\textrm{max}}$ for which the streaming flow
topology for a circular cylinder ($\textrm{AR} = 1$) lies in the finite thickness 
DC layer regime (figure \ref{fig:vell_hetero}(\textit{d})). 
We focus on the heteroclinic orbits highlighted in figure
\ref{fig:vell_hetero}(\textit{d}). Recalling our observations from the lattice system, we 
predict that introducing fore-aft shape asymmetry will lead to the
breaking of these orbits, as captured by the heteroclinic orbit bifurcation from Phase
\rom{2} $\to$ \rom{5} (figure \ref{fig:vell_hetero}(\textit{b, c})) in the main text.   
Indeed, testing a shape with curvature asymmetry (i.e. $\textrm{AR} > 1$) confirms this
prediction, as seen from the flow topology in figure \ref{fig:vell_hetero}(\textit{e}).

After testing our understanding against the effects of breaking symmetry, we
analyse how flow topology evolves on further increase in curvature variation 
(i.e. increase in $\textrm{AR}$) for the body. Figure \ref{fig:vell_sadcent}(\textit{e}) depicts
the streaming flow topology for the body with $\textrm{AR} > 1$, where the shape
parameters and flow topology are retained from figure \ref{fig:vell_hetero}(\textit{e}). 
We focus on the highlighted
saddle (defining DC layer extent) forming the homoclinic orbit, and the centre within 
this orbit. As $\textrm{AR}$ increases, (figure \ref{fig:vell_sadcent}(\textit{a})), the
local curvature on the body near the saddle increases. 
Based on the transitions \rom{5} $\to$ \rom{7} and \rom{3} $\to$ \rom{6}
associated with such a curvature increase in the lattice phase space, 
we predict that this will lead to a decrease in the saddle--centre distance, while
shrinking the corresponding homoclinic orbit. Then beyond a critical value of
$\textrm{AR}$ the orbit will eventually disappear through a saddle--centre annihilation,
similar to the bifurcation seen from Phase \rom{5} $\to$ \rom{7}.
Figure \ref{fig:vell_sadcent}(\textit{b, c, d}) demonstrate the reduced Hamiltonian form contours for this
system and showcase the saddle--centre bifurcation. 
Indeed, testing a shape with curvature variation (higher $\textrm{AR}$) confirms these
predictions, as seen from the flow topologies in figure
\ref{fig:vell_sadcent}(\textit{f, g}). 

\begin{figure}[h!]
    \centerline{\includegraphics[width=0.9\textwidth]{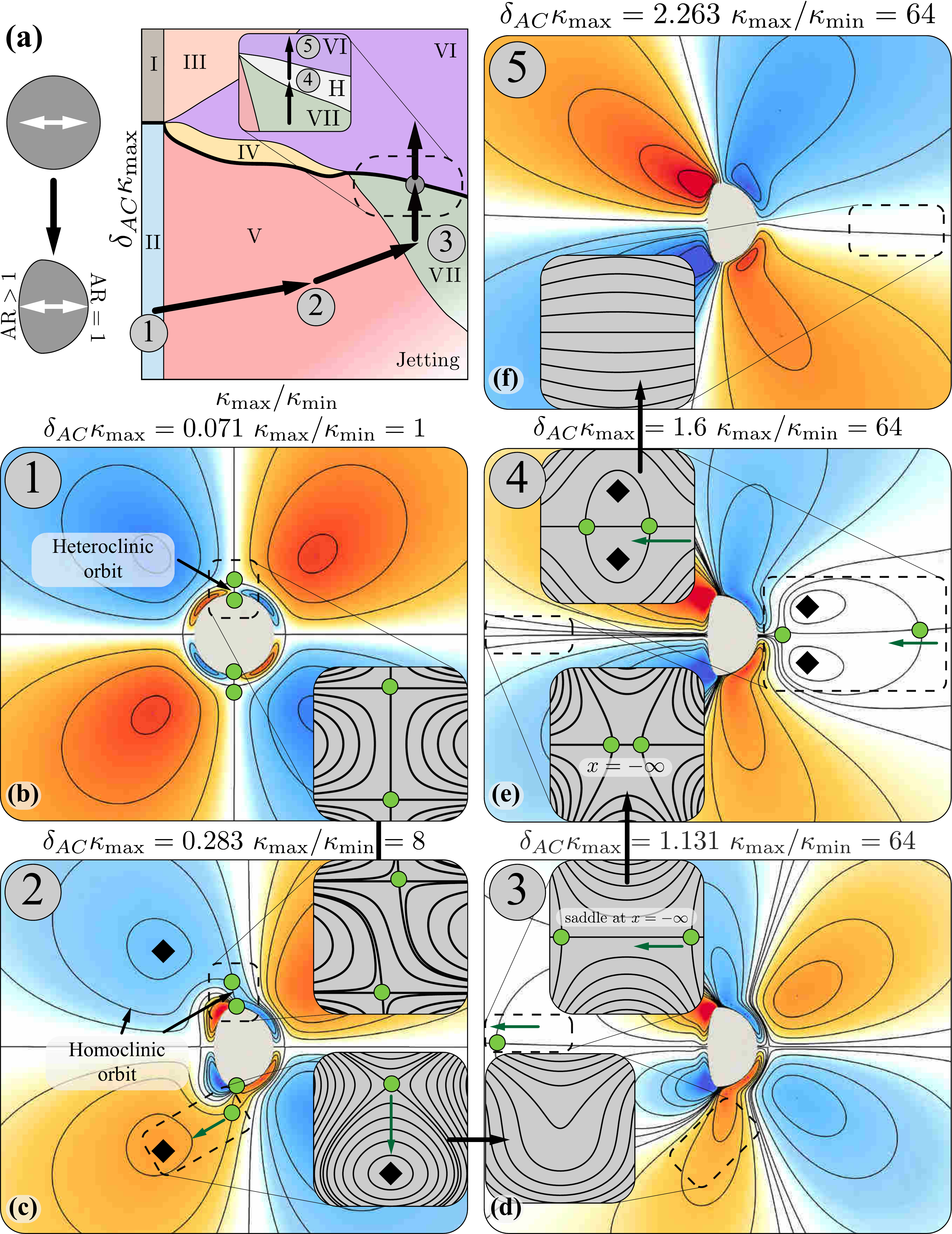}}
    \caption{Flow topology manipulation. (\textit{a}) Illustration of morphing a circular
        cylinder into a hybrid circular--elliptic cylinder. Mapping 
        of the observed transitions on the lattice phase space. (\textit{b-f}) present the 
        different topologies observed on geometric and background flow variation, with 
        the concerned critical points highlighted and the predictions (reduced 
        Hamiltonian form contours) illustrated as grey contours.}\label{fig:vell_ps}
\end{figure}

For completion, figure \ref{fig:vell_ps} presents a step-by-step systematic
variation of geometric and background flow properties for the morphing from a circular
cylinder to a hybrid circular--elliptic cylinder. We observe a wide range of flow
topologies, which are predicted (reduced Hamiltonian as grey contours) by drawing 
intuition from the transitions observed in the lattice phase space with corresponding 
confirmations from simulations. We note that the same set of bifurcations as that of the
circular--square cylinder in the main text, are encountered for the circular--elliptic cylinder.
Figure \ref{fig:vell_ps}(\textit{b, c}) illustrate the
heteroclinic orbit bifurcation (prediction and  confirmation), observed on breaking 
symmetry described above. Figure \ref{fig:vell_ps}(\textit{c, d}) illustrate the
saddle--centre bifurcation (prediction and confirmation), observed on curvature 
variation described above.  
Figure \ref{fig:vell_ps}(\textit{d, e}) and figure \ref{fig:vell_ps}(\textit{e, f}) 
illustrate the elliptic reflecting umbilic
bifurcation and the hyperbolic reflecting umbilic bifurcation, respectively, 
observed on increasing $\delta_{AC}$ in the lattice phase space, with the predicted flow topology
change based on the reduced Hamiltonian contours and confirmations against flow
topologies observed in simulations.

In summary, this section illustrates an additional example of how our approach can be 
employed to predict and design streaming flow topologies, in a rational fashion.

\vspace{-10pt}
\subsection{Varying ellipse aspect ratio}
\vspace{-10pt}

\begin{figure}[h!]
    \centerline{\includegraphics[width=\textwidth]{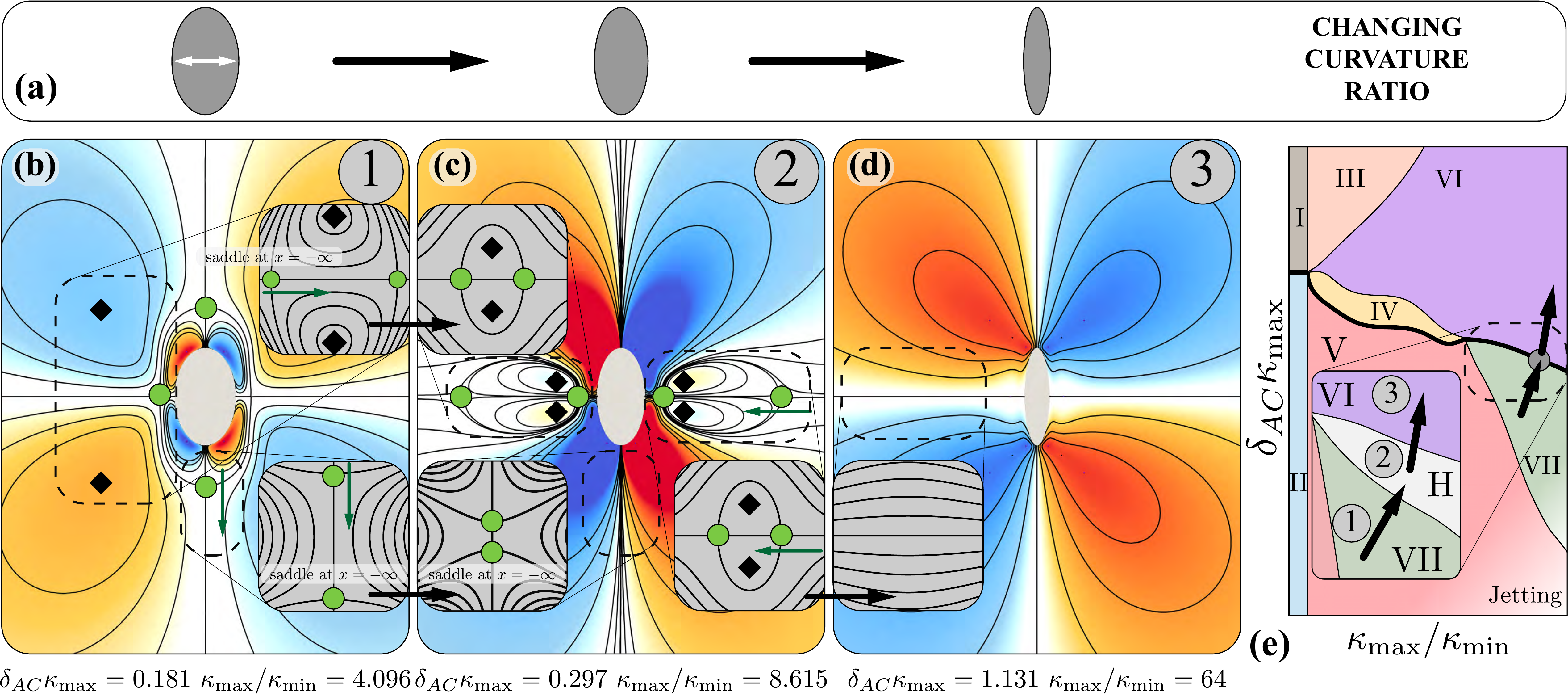}}
    \caption{Flow topology manipulation. (\textit{a}) Illustration of varying curvature of
        an ellipse by varying the aspect ratio $\textrm{AR}$. (\textit{b-d}) present the 
        different topologies observed on geometric variation, with 
        the concerned critical points highlighted and the predictions (reduced 
        Hamiltonian form contours) illustrated as grey contours. (\textit{e}) Mapping 
    of the observed transitions on the lattice phase space.}\label{fig:sinell}
\end{figure}

Figure \ref{fig:sinell}(\textit{a}) illustrates the use of our phase space in the case of
ellipses of varying aspect ratio (defined as height/breadth). We keep the major axis (height = $L$) fixed. 
This allows us to introduce curvature variation in a regulated fashion by spanning AR. 
We start by choosing a $\delta_{AC} \kappa_{\textrm{max}}$ and $\textrm{AR}$ for which 
the streaming flow topology for the ellipse lies in the finite thickness 
DC layer regime (figure \ref{fig:sinell}(\textit{b})). We note that a mapping of this
local flow topology to Phase \rom{1} and \rom{2} of our lattice system is not possible due
to the variation of curvature on an ellipse (Phase \rom{1} and \rom{2} necessitate constant 
curvature). Additionally, mapping to Phases \rom{3}, \rom{4} and \rom{5} does not exist as
we do not break fore-aft or top-down shape symmetry when we vary AR for an ellispe (figure
\ref{fig:sinell}(\textit{a})). This means that we don't break symmetry for the DC layer
bounding heteroclinic orbits which are necessary for Phases \rom{3}, \rom{4} and \rom{5} to exist.

We focus on the highlighted saddles bounding the DC layers on the vertical axis, near the
top and bottom ends of the ellipse. 
These structures closely resemble Phase \rom{7} of figure 7(\textit{k, n}) in the main text,
where the second saddle (not imaged in experiments) is located at infinity.
Thus, as $\delta_{AC} \kappa_{\textrm{max}}$ increases (on increasing AR), we predict that the 
saddles will progressively move outwards, to approach the
saddles at infinity and undergo an elliptic reflecting umbilic bifurcation. This has the 
overall effect to ``open up" the DC layers on the vertical axis of the ellipse. 
Additionally, we note that the highlighted flow structures on the left/right of the streaming shape can 
be mapped to Phase H of figure 7(d, g) (with one saddle at infinity). Therefore, as a side
effect of the increase in $\delta_{AC}\kappa_{\textrm{max}}$, we predict that the saddles at infinity enter the imaged 
domain and form closed recirculating regions on the left and right side of the ellipse.
Computations of figure \ref{fig:sinell}(\textit{c}) confirm this intuition.

We then focus on the highlighted saddles and centres forming recirculating regions on the
horizontal axis of the ellipse in figure \ref{fig:sinell}(\textit{c}).
These structures closely resemble the hidden Phase H of figure 7(\textit{d, g}).
Based on our phase space, we predict that as $\delta_{AC} \kappa_{\textrm{max}}$ increases
(on increasing AR), the system will transition 
to a new topology corresponding to Phase \rom{6}, via a hyperbolic 
reflecting umbilic bifurcation. This is a consequence of the saddles and centres moving 
closer and closer, eventually collapsing and vanishing.
Once again, computations of figure \ref{fig:sinell}(\textit{d}) confirm this intuition.

We note that in the ellipse case, due to its geometrical properties, some phases captured
in our lattice phase space do not exist (see above discussion). Nonetheless, the phases and 
the transitions that do exist remain consistent with the lattice phase space, and so our
analysis is still valid.

\bibliographystyle{unsrtnat}
\bibliography{arxiv.bbl}

\end{document}